\documentclass[preprint2]{proto}
\usepackage{times}
\newcommand{\refs}{\par\noindent\hangindent=1pc\hangafter=1}
\newcommand{\LX}{L_{\rm X}}
\newcommand{\Mdot}{\dot M}

\newcommand{\Msunperyr}{M_{\odot}\,{\rm yr}^{-1}}

\newcommand{\percc}{\rm \,cm^{-3}}
\newcommand{\psqcm}{{\rm cm}^{-2}}
\newcommand{\persqcm}{\rm \,cm^{-2}}
\newcommand{\gpersqcm}{\rm \,g\,cm^{-2}}

\newcommand{\ps}{{\rm s}^{-1}}

\newcommand{\kmps}{{\rm km}\,{\rm s}^{-1}}
\newcommand{\kms}{{\rm km}\,{\rm s}^{-1}}
\newcommand{\erg}{{\rm erg}}

\def\micron{\hbox{$\mu$m}}
\newcommand{\Htwo}{H$_2$}
\newcommand{\Lya}{Ly\,$\alpha$}
\newcommand{\be}{\begin{equation}}
\newcommand{\ee}{\end{equation}}

\begin{document}

\title{\textbf{\LARGE Gaseous Inner Disks}}

\author {\textbf{\large Joan R. Najita}} 
\affil{\small\em National Optical Astronomy Observatory}

\author {\textbf{\large John S. Carr}}
\affil{\small\em Naval Research Laboratory}

\author {\textbf{\large Alfred E. Glassgold}}
\affil{\small\em University of California, Berkeley}

\author {\textbf{\large Jeff A. Valenti}}
\affil{\small\em Space Telescope Science Institute}

\begin{abstract}
\baselineskip = 11pt
\leftskip = 0.65in 
\rightskip = 0.65in
\parindent=1pc
{\small As the likely birthplaces of planets and an essential conduit 
for the buildup of stellar masses, inner disks are of fundamental 
interest in star and planet formation.  
Studies of the gaseous component of inner disks are of interest 
because of their ability to probe 
the dynamics, physical and chemical structure, and gas content of this region. 
We review the observational and theoretical developments in this field, 
highlighting the potential of such studies to, e.g., 
measure inner disk truncation radii, 
probe the nature of the disk accretion process, and 
chart the evolution in the gas content of disks.  
Measurements of this kind have the potential to 
provide unique insights on the physical processes governing 
star and planet formation.  
\\~\\~\\~}%leave this in to get the correct vertical space after the abstract

%\end{list}
\end{abstract}

\section{\textbf{INTRODUCTION}}

Circumstellar disks play a fundamental role in the formation of 
stars and planets. 
A significant fraction of the mass of a star is thought to be 
built up by accretion through the disk.
The gas and dust in the inner disk ($r<$10\,AU) 
also constitute the likely material from which planets form.  
As a result, observations of the gaseous 
component of inner disks have the potential to provide critical clues to 
the physical processes governing star and planet formation.

From the planet formation perspective, probing the structure,
gas content, and dynamics of inner disks is of interest, since
they all play important
roles in establishing the architectures of planetary systems
(i.e., planetary masses, orbital radii, and eccentricities).
For example, the lifetime of gas in the inner disk
(limited by accretion onto the star, photoevaporation, 
and other processes)
places an upper limit on the timescale for giant planet formation
(e.g., {\em Zuckerman et al.}, 1995).

The evolution of gaseous inner disks may also bear 
on the efficiency of 
orbital migration and the eccentricity evolution of giant and terrestrial
planets.  Significant inward orbital migration, induced
by the interaction of planets with a gaseous disk, is 
implied by the small orbital radii of extrasolar
giant planets compared to their likely formation distances
(e.g., {\em Ida and Lin}, 2004).  
The spread in the orbital radii of the planets (0.05--5\,AU)
has been further taken to indicate that the timing of the dissipation 
of the inner disk sets the final orbital radius of the planet
({\em Trilling et al.}, 2002).
Thus, understanding 
how inner disks dissipate may impact 
our understanding of the origin of planetary orbital radii.
Similarly, residual gas in the terrestrial planet region may play 
a role in defining the final masses and eccentricities of 
terrestrial planets.  Such issues have a strong connection
to the question of the likelihood of solar systems like our own.

An important issue from the perspective of both star and planet
formation is the nature of the physical mechanism that is 
responsible for disk accretion.  
Among the proposed mechanisms, 
perhaps the foremost is the magnetorotational instability 
({\em Balbus and Hawley}, 1991) although other possibilities exist.   
Despite the significant theoretical progress that has been made 
in identifying plausible accretion mechanisms 
(e.g., {\em Stone et al.}, 2000), there is little observational 
evidence that any of these processes are active in disks. 
Studies of the gas in inner disks offer opportunities 
to probe the nature of the accretion process. 

For these reasons, it is of interest to probe 
the dynamical state, physical and chemical structure, and the 
evolution of the gas content of inner disks.
We begin this Chapter with a brief review of the 
development of this field and an overview of 
how high resolution spectroscopy can be used to study the 
properties of inner disks (Section 1). 
Previous reviews provide additional background on these topics 
(e.g., {\em Najita et al.}, 2000).  
In Sections 2 and 3, we review recent observational and theoretical 
developments in this field, first describing 
observational work to date on the gas in 
inner disks, and then describing theoretical models 
for the surface and interior regions of disks. 
In Section 4, we look to the future, highlighting several topics 
that can be explored using the tools discussed in Sections 2 and 3.

\bigskip
\noindent
\textbf{1.1 Historical Perspective} 
\bigskip

One of the earliest studies of gaseous inner disks was 
the work by Kenyon and Hartmann on FU Orionis objects. 
They showed that many of the peculiarities of these systems 
could be explained in terms of 
an accretion outburst in a disk surrounding 
a low-mass young stellar object (YSO; cf.\ {\em Hartmann and Kenyon},\ 1996).
In particular, the varying spectral type of FU Ori objects in 
optical to near-infrared spectra, evidence for double-peaked 
absorption line profiles, and the decreasing widths of absorption 
lines from the optical to the near-infrared argued for an origin 
in an optically thick gaseous atmosphere in the inner region of 
a rotating disk.
Around the same time, 
observations of CO vibrational overtone emission, first in the BN object
({\em Scoville et al.},\ 1983) and later in other high and low mass objects
({\em Thompson}, 1985; {\em Geballe and Persson}, 1987; 
{\em Carr}, 1989), revealed the 
existence of hot, dense molecular gas plausibly located in a disk.
One of the first models for the CO overtone emission ({\em Carr}, 1989) placed
the emitting gas in an optically-thin inner region of an accretion
disk. 
However, only the observations of the BN object had sufficient 
spectral resolution to constrain the kinematics of the emitting gas. 

The circumstances under which a disk would produce emission or 
absorption lines of this kind 
were explored in early models
of the atmospheres of gaseous accretion disks under the influence
of external irradiation (e.g., {\em Calvet et al.},\ 1991).  The models
interpreted the FU Ori absorption features 
as a consequence of midplane accretion rates high enough to 
overwhelm external irradiation in establishing a temperature
profile that decreases with disk height.  At lower accretion rates,
the external irradiation of the disk was expected to induce a
temperature inversion in the disk atmosphere, producing emission
rather than absorption features from the disk atmosphere.  
Thus the models potentially provided an explanation for the 
FU Ori absorption features and CO emission lines that had been 
detected.

By PPIV ({\em Najita et al.},\ 2000), high-resolution spectroscopy 
had demonstrated that CO overtone emission shows the dynamical signature of a
rotating disk ({\em Carr et al.},\ 1993; {\em Chandler et al.},\ 1993), thus
confirming theoretical expectations and opening the door
to the detailed study of gaseous inner disks in a larger number of
YSOs.
The detection of CO fundamental emission (Section 2.3) and emission
lines of hot H$_2$O (Section 2.2) had also added new probes of the 
inner disk gas.

Seven years later, at PPV, we find both a growing
number of diagnostics available to probe gaseous inner disks as
well as increasingly detailed information that can be gleaned from
these diagnostics.  Disk diagnostics familiar from PPIV 
have been used to infer the intrinsic line broadening of disk
gas, possibly indicating evidence for turbulence in disks (Section 2.1).  
They also demonstrate the differential rotation of disks,  
provide evidence for non-equilibrium molecular abundances (Section 2.2),  
probe the inner radii of gaseous disks (Section 2.3), 
and are being used to probe the gas dissipation timescale 
in the terrestrial planet region (Section 4.1).
Along with these developments, new spectral line diagnostics have 
been used as probes of the gas in inner disks.  These 
include transitions of molecular hydrogen at UV, near-infrared, and  
mid-infrared wavelengths (Sections 2.4, 2.5) and the fundamental ro-vibrational 
transitions of the OH molecule (Section 2.2).  Additional potential 
diagnostics are discussed in Section 2.6.

\bigskip
\noindent
\textbf{1.2 High Resolution Spectroscopy of Inner Disks}
\bigskip
%\subsection{High Resolution Spectroscopy of Inner Disks}

The growing suite of diagnostics can be used to probe 
inner disks using standard high resolution spectroscopic 
techniques. 
Although inner disks are typically too small to resolve 
spatially at the distance of the nearest star forming regions,
we can utilize the likely differential rotation 
of the disk along with high spectral resolution to separate 
disk radii in velocity.  
At the warm temperatures ($\sim$100\,K--5000\,K) and high densities of inner
disks, molecules are expected to be abundant in the gas phase and
sufficiently excited to produce rovibrational features in the
infrared.  Complementary atomic transitions are likely to be good 
probes of the hot inner disk and the photodissociated surface 
layers at larger radii. 
By measuring multiple transitions of different species, 
we should therefore be able
to probe the temperatures, column densities, and abundances
of gaseous disks as a function of radius.

With high spectral 
resolution we can resolve individual lines, which facilitates 
the detection of weak spectral features. 
We can also work around  
telluric absorption features, using the radial velocity of the 
source to shift its spectral features out of telluric 
absorption cores. 
This approach makes it 
possible to study a variety of atomic and molecular 
species, including those present in the Earth's atmosphere.

Gaseous spectral features are expected in a variety of situations.  
As already mentioned, significant vertical variation in the temperature 
of the disk atmosphere will produce emission (absorption) features if the 
temperature increases (decreases) with height ({\em Calvet et al.}, 1991; 
{\em Malbet and Bertout}, 1991).  
In the general case, when the disk is optically thick, observed spectral
features measure only the atmosphere of the disk and are unable to 
probe directly the entire disk column density, a situation familiar 
from the study of stellar atmospheres.

Gaseous emission features are also expected 
from regions of the disk that are optically thin in the continuum. 
Such regions might arise as a result of dust sublimation 
(e.g., {\em Carr}, 1989) or as a consequence of grain growth 
and planetesimal formation. 
In these scenarios, the disk would have a low continuum opacity 
despite a potentially large gas column density. 
Optically thin regions can also be produced by a significant 
reduction in the total column density of the disk.  This situation 
might occur as a consequence of giant planet formation, in which 
the orbiting giant planet carves out a ``gap'' in the disk.  
Low column densities would also be characteristic of a dissipating disk. 
Thus, we should be able to use gaseous emission lines to probe 
the properties of inner disks in a variety of interesting 
evolutionary phases.

\section{\textbf{OBSERVATIONS OF GASEOUS INNER DISKS}}

\bigskip
\noindent
\textbf{2.1 CO Overtone Emission}
\bigskip
%\subsection{CO Overtone Emission}

The CO molecule is expected to be 
abundant in the gas phase over a wide range of 
temperatures, from the temperature at which it condenses on grains 
($\sim$20\,K) up to its thermal dissociation temperature 
($\sim$4000\,K at the densities of inner disks). 
As a result, CO transitions are expected to probe 
disks from their cool outer reaches ($>$100\,AU) 
to their innermost radii.
Among these, the overtone transitions of CO ($\Delta v$=2, 
$\lambda$=2.3$\mu$m) were the emission line diagnostics first 
recognized to probe the gaseous inner disk.

CO overtone emission is detected in both low and high mass young
stellar objects, but only in a small fraction of the objects observed.
It appears more commonly among higher luminosity objects. Among 
the lower luminosity stars, it is detected from embedded protostars 
or sources with energetic outflows
({\em Geballe and Persson}, 1987; {\em Carr}, 1989; 
{\em Greene and Lada}, 1996; {\em Hanson et al.}, 1997;
{\em Luhman et al.}, 1998; {\em Ishii et al.}, 2001; 
{\em Figueredo et al.}, 2002; {\em Doppmann et al.}, 2005).
The conditions required to excite the overtone emission,  
warm temperatures ($\gtrsim 2000$ K) and high densities 
($>$$10^{10}\percc$), may be met in disks
({\em Scoville et al.}, 1983; {\em Carr}, 1989; 
{\em Calvet et al.}, 1991), inner winds ({\em Carr}, 1989), 
or funnel flows ({\em Martin}, 1997). 

High resolution spectroscopy can be used to distinguish among these 
possibilities.  The observations typically find strong 
evidence for the disk interpretation.  The emission line profiles 
of the $v$=2--0 bandhead in most cases 
show the characteristic signature of bandhead emission from
symmetric, double-peaked line profiles originating in 
a rotating disk (e.g., {\em Carr et al.}, 1993; {\em Chandler et al.}, 1993; 
{\em Najita et al.}, 1996; {\em Blum et al.}, 2004).  The symmetry of the observed 
line profiles argues against the likelihood that the emission arises 
in a wind or funnel flow, since inflowing or outflowing gas is 
expected to produce line profiles with  red- or blue-shifted 
absorption components (alternatively line asymmetries) of the 
kind that are seen in the hydrogen Balmer lines of T Tauri stars 
(TTS).
Thus high resolution spectra provide strong evidence for 
rotating inner disks. 

The velocity profiles of the CO overtone emission are normally
very broad ($>$100\,$\kmps$).  In lower mass stars 
($\sim$$1 M_\odot$), the emission profiles show that the emission 
extends from very close to the star,
$\sim$0.05\,AU, out to $\sim$0.3\,AU
(e.g., {\em Chandler et al.}, 1993; {\em Najita et al.}, 2000).
The small radii are consistent with the high excitation
temperatures measured for the emission ($\sim$1500--4000\,K).
Velocity resolved spectra have also been modeled in a number
of high mass stars ({\em Blum et al.}, 2004; {\em Bik and Thi}, 2004), 
where the CO emission is found to arise at radii $\sim 3$\,AU.

The large near-infrared excesses of the sources in which CO 
overtone emission is detected imply that the warm emitting gas 
is located in a vertical temperature inversion region in the disk atmosphere. 
Possible heating sources for the temperature inversion include: 
external irradiation by the star at optical through UV wavelengths 
(e.g., {\em Calvet et al.}, 1991; {\em D'Alessio et al.}, 1998) 
or by stellar X-rays ({\em Glassgold et al.}, 2004; henceforth GNI04); 
turbulent heating in the disk atmosphere generated by a 
stellar wind flowing over the disk surface ({\em Carr et al.}, 1993);  
or the dissipation of turbulence generated by disk accretion 
(GNI04). 
Detailed predictions of how these mechanisms heat the gaseous 
atmosphere are needed in order to 
use the observed bandhead emission strengths and profiles to 
investigate the origin of the temperature inversion. 

The overtone emission provides an additional clue 
that suggests a role for turbulent dissipation 
in heating disk atmospheres.  
Since the CO overtone bandhead is made up of closely spaced lines 
with varying inter-line spacing and optical depth, the emission 
is sensitive to the intrinsic line broadening of the emitting 
gas (as long as the gas is not optically thin).  
It is therefore possible to distinguish intrinsic line 
broadening from macroscopic motions such as rotation.  
In this way, 
one can deduce from spectral synthesis modeling 
that the lines are suprathermally broadened, with line widths 
approximately Mach 2 ({\em Carr et al.}, 2004; {\em Najita et al.}, 1996). 
{\em Hartmann et al.}\ (2004) find further evidence for 
turbulent motions in disks based on high resolution 
spectroscopy of CO overtone absorption in FU Ori objects.

Thus disk atmospheres appear to be turbulent.  The turbulence 
may arise as a consequence of turbulent angular momentum 
transport in disks, as in the magnetorotational instability 
(MRI; {\em Balbus and Hawley}, 1991) or the global baroclinic instability 
({\em Klahr and Bodenheimer}, 2003).  Turbulence in the upper disk 
atmosphere may also be generated by a wind blowing over the 
disk surface.

\bigskip
\noindent
\textbf{2.2 Hot Water and OH Fundamental Emission}
\bigskip
%\subsection {Hot Water and OH Fundamental Emission}

Water molecules are also expected to be abundant in disks over 
a range of disk radii, from the temperature at which water 
condenses on grains ($\sim$150 K) up to its thermal dissociation 
temperature ($\sim$2500 K).  Like the CO overtone 
transitions, the rovibrational transitions of water are also 
expected to probe the high density conditions in disks. 
While the strong telluric absorption produced by water vapor in the
Earth's atmosphere will restrict the study of cool water to
space or airborne platforms, it is possible to observe from the
ground water that is much hotter than the Earth's atmosphere.
Very strong emission from hot water can be detected in the near-infrared 
even at low spectral resolution (e.g., SVS-13; {\em Carr et al.}, 2004).  
More typically, high resolution spectroscopy of
individual lines is required to detect much weaker emission lines.

For example, emission from individual lines of water in the
$K$- and $L$-bands have been detected in a few stars (both low 
and high mass) that also show CO overtone emission 
({\em Carr et al.}, 2004; {\em Najita et al.}, 2000; {\em Thi and Bik}, 2005).
Velocity resolved spectra show that the widths of the
water lines are consistently narrower than those of the 
CO emission lines.  Spectral synthesis modeling further shows 
that the excitation temperature of the water emission
(typically $\sim$1500\,K),
is less than that of the CO emission.  These results
are consistent with both the water and CO originating in a
differentially rotating
disk with an outwardly decreasing temperature profile.  That is,
given the lower dissociation temperature of water ($\sim$2500\,K)
compared to CO ($\sim$4000\,K),
CO is expected to extend inward to smaller radii than 
water, i.e., to higher velocities and temperatures. 

The $\Delta v$=1 OH fundamental transitions 
at 3.6$\mu$m have also been detected in the spectra of two actively 
accreting sources, SVS-13 and V1331~Cyg, that also show CO 
overtone and hot water emission  (Carr et al., in preparation).  
As shown in Fig.~1, these 
features arise in a region that is crowded with spectral 
lines of water and perhaps other species.  Determining the strengths of 
the OH lines will, therefore, require making corrections for 
spectral features that overlap closely in wavelength. 

Spectral synthesis modeling of the detected CO, H$_2$O and OH features 
reveals relative abundances 
that depart significantly from chemical equilibrium 
(cf.\ {\em Prinn}, 1993), with the 
relative abundances of H$_2$O and OH a factor of 2--10 below that of CO 
in the region of the disk probed by both diagnostics 
({\em Carr et al.}, 2004; Carr et al., in preparation; 
see also {\em Thi and Bik}, 2005).
These abundance ratios may arise from strong vertical abundance 
gradients produced by the external irradiation of the disk 
(see Section 3.4). 

\begin{figure}[b!]
\plotfiddle{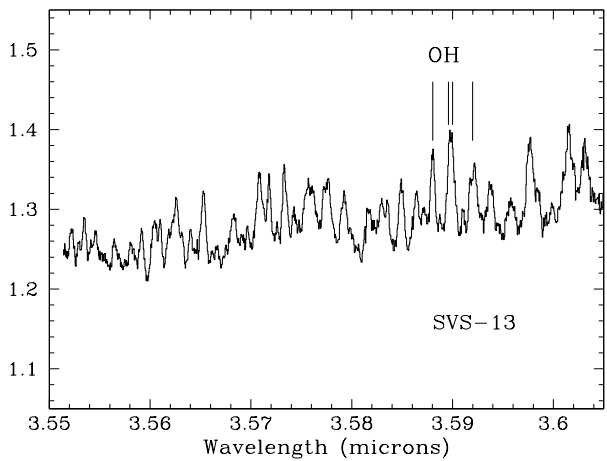}{0.05in}{270}{180}{250}{-8}{0}
\vspace{-.7in}
\caption{\small\baselineskip 10pt
OH fundamental ro-vibrational emission from SVS-13  
on a relative flux scale.}
\end{figure}

\bigskip
\noindent
\textbf{2.3 CO Fundamental Emission}
\bigskip
%\subsection{CO Fundamental Emission}

The fundamental ($\Delta v$=1) transitions of CO at 4.6$\mu$m 
are an important probe of inner disk gas in part because of 
their broader applicability compared, e.g., to the CO overtone lines. 
As a result of their comparatively small A-values, 
the CO overtone transitions require large column densities of warm gas 
(typically in a disk temperature inversion region) in order to produce 
detectable emission.  
Such large column densities of warm gas may be rare 
except in sources with the largest accretion rates, i.e., 
those best able to tap a large accretion energy budget and 
heat a large column density of the disk atmosphere. 
In contrast, the CO fundamental transitions, with their much larger 
A-values, should be detectable 
in systems with more modest column densities of warm gas, 
i.e., in a broader range of sources.   
This is borne out in high resolution spectroscopic surveys for CO 
fundamental emission from TTS ({\em Najita et al.}, 2003) 
and Herbig AeBe stars ({\em Blake and Boogert}, 2004) 
which detect emission from 
essentially all sources with accretion rates typical of these 
classes of objects.

\begin{figure}[t!]
\vspace{-1.6in}
\plotfiddle{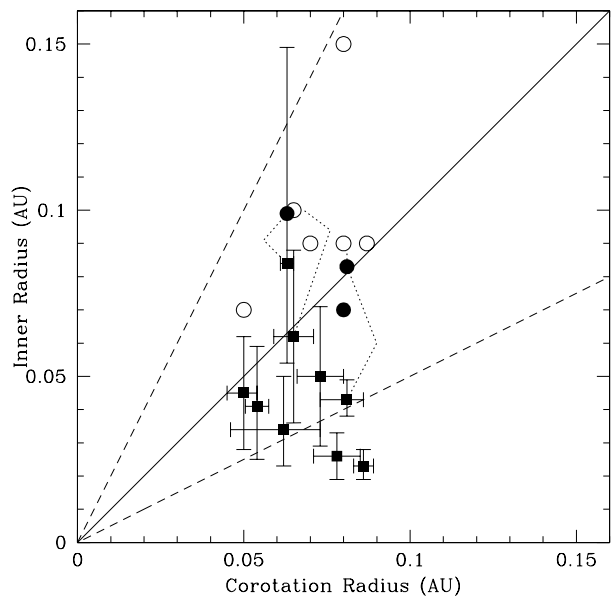}{1.50in}{0.}{240}{240}{5}{150}
\vspace{-0.3in}
\caption{\small\baselineskip 10pt
Gaseous inner disk radii for TTS from CO fundamental emission 
(filled squares) compared with corotation radii for the same sources. 
Also shown are dust inner radii from near-infrared interferometry 
(filled circles; {\em Akeson et al.}, 2005a,b) or spectral energy 
distributions 
(open circles; {\em Muzerolle et al.}, 2003).  
The solid and dashed lines indicate an inner radius equal to, 
twice, and 1/2 the corotation radius.
The points for the three stars with measured inner radii
for both the gas and dust are connected by dotted lines.
Gas is observed to extend inward of the dust inner radius 
and typically inward of the corotation radius.
} 
\end{figure}

In addition, the lower temperatures
required to excite the CO $v$=1--0 transitions make these transitions 
sensitive to cooler gas at larger disk radii, beyond the 
region probed by the 
CO overtone lines.  Indeed, the measured line profiles for the 
CO fundamental emission are broad (typically 50--100\,$\kmps$ FWHM) and 
centrally peaked, in contrast to the CO overtone lines which are 
typically double-peaked.  These velocity profiles suggest 
that the CO fundamental emission arises from a wide range of radii, 
from $\lesssim$0.1\,AU out to 1--2\,AU in disks 
around low mass stars, i.e., the terrestrial planet region of 
the disk ({\em Najita et al.}, 2003). 

CO fundamental emission spectra typically show symmetric emission 
lines from multiple vibrational states (e.g., $v$=1--0, 2--1, 3--2); 
lines of $^{13}$CO can also be detected when the emission is strong 
and optically thick. 
The ability to study multiple vibrational states as well as 
isotopic species within a limited spectral range makes the CO 
fundamental lines an appealing choice to probe gas in the inner disk 
over a range of temperatures and column densities.  
The relative strengths of the lines 
also provide insight into the excitation mechanism for the emission.

In one source, the Herbig AeBe star HD141569, the excitation
temperature of the rotational levels ($\sim$200\,K) is much lower
than the excitation temperature of the vibrational levels ($v$=6
is populated), which is suggestive of UV pumping of cold gas 
({\em Brittain et al.}, 2003).  
The emission lines from the source are narrow, indicating
an origin at $\gtrsim$17\,AU.  The lack of fluorescent emission
from smaller radii strongly suggests that the region within 17\,AU 
is depleted of gaseous CO.  
Thus detailed models of the fluorescence process can be used
to constrain the gas content in the inner disk region 
(S. Brittain, personal communication).

Thus far HD141569 appears to be an unusual case.  For the majority 
of sources from which CO fundamental is detected, the relative line 
strengths are consistent with emission from thermally excited
gas.  They indicate typical excitation temperatures of 
1000--1500\,K and CO column densities of $\sim$$10^{18}\persqcm$  
for low mass stars. 
These temperatures are much warmer than the dust temperatures at
the same radii implied by spectral energy distributions (SEDs) 
and the expectations of some disk atmosphere models
(e.g., {\em D'Alessio et al.}, 1998).  
The temperature difference can be 
accounted for by   
disk atmosphere models 
that allow for the thermal decoupling of the
gas and dust (Section 3.2).

For CTTS systems in which the inclination is known, we can convert 
a measured HWZI velocity for the emission to an inner radius.  The CO inner 
radii, thus derived, are typically $\sim$0.04\,AU for TTS 
({\em Najita et al.}, 2003; {\em Carr et al.}, in preparation), 
smaller than the inner radii that are measured for the 
dust component either through interferometry 
(e.g., {\em Eisner et al.}, 2005; {\em Akeson et al.}, 2005a; 
{\em Colavita et al.}, 2003; see chapter by {\em Millan-Gabet et al.})\ or 
through the interpretation 
of SEDs (e.g., {\em Muzerolle et al.}, 2003).
This shows that gaseous disks extend inward 
to smaller radii than dust disks, a result that is not 
surprising given the relatively low sublimation temperature 
of dust grains ($\sim$1500--2000 K) compared to the 
CO dissociation temperature ($\sim$4000 K). 
These results are consistent with the suggestion that 
the inner radius of the dust disk is defined by 
dust sublimation rather than by physical truncation
({\em Muzerolle et al.}, 2003; {\em Eisner et al.}, 2005).

Perhaps more interestingly, the inner radius of the CO emission 
appears to extend up to and usually within the corotation 
radius (i.e., the radius at which the disk rotates at the same 
angular velocity as the star; Fig.~2).  In the current paradigm for 
TTS, a strong stellar magnetic field 
truncates the disk near the corotation radius. 
The coupling between the stellar magnetic field and the gaseous inner disk 
regulates the rotation of the star, bringing 
the star into corotation with the disk at the coupling radius. 
From this region emerge both energetic (X-)winds and magnetospheric 
accretion flows (funnel flows; {\em Shu et al.}, 1994).  
The velocity extent of the CO fundamental emission shows that gaseous
circumstellar disks indeed extend inward beyond the dust destruction
radius to the corotation radius (and beyond), 
providing the material that feeds both X-winds and funnel flows.
Such small coupling radii are consistent with the rotational rates 
of young stars.

\begin{figure}[t!]
\vspace{-0.1in}
\plotfiddle{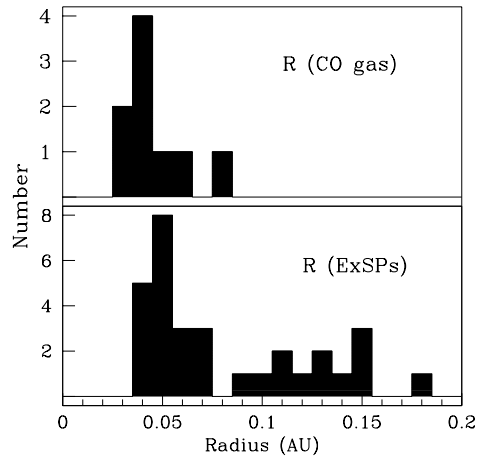}{0.05in}{0.}{180}{180}{20}{0}
\vspace{-0.1in}
\caption{\small\baselineskip 10pt
The distribution of gaseous inner radii, measured with 
the CO fundamental transitions, compared to the 
distribution of orbital radii of short-period
extrasolar planets.
A minimum planetary orbital radius of $\sim$0.04\,AU 
is similar to the minimum gaseous inner radius
inferred from the CO emission line profiles.
} 
\end{figure}

It is also interesting to compare the distribution of inner
radii for the CO emission with the orbital radii of
the ``close-in'' extrasolar giant planets (Fig.\ 3).
Extra-solar planets discovered by radial velocity surveys
are known to pile-up near a minimum radius of 0.04 AU.
The similarity between these distributions is roughly consistent 
with the idea that the truncation of the inner disk 
can halt the inward orbital migration of a giant 
planet ({\em Lin et al.}, 1996).  In detail, however, the planet 
is expected to migrate slightly inward 
of the truncation radius, to the 2:1 resonance, 
an effect that is not seen in the present 
data.  A possible caveat is that the wings of the CO lines may 
not trace Keplerian motion or that the innermost gas is not 
dynamically significant.  It would be interesting to explore 
this issue further since the results impact our understanding 
of planet formation and the origin of planetary architectures.  
In particular, the existence of a stopping mechanism implies 
a lower efficiency for giant planet formation, e.g., compared 
to a scenario in which multiple generations of planets form and 
only the last generation survives  
(e.g., {\em Trilling et al.}, 2002).

\bigskip
\noindent
\textbf{2.4 UV Transitions of Molecular Hydrogen}
\bigskip
%\subsection{UV Transitions of Molecular Hydrogen} 

Among the diagnostics of inner disk gas 
developed since PPIV, perhaps the most interesting 
are those of 
\Htwo.
\Htwo\ is presumably the dominant gaseous species in disks, due
to high elemental abundance, low depletion onto grains, and robustness
against dissociation. Despite its expected ubiquity, \Htwo\ is difficult
to detect because permitted electronic transitions are in the far
ultraviolet (FUV) and accessible only from space. Optical and 
rovibrational IR
transitions have radiative rates that are 14 orders of magnitude
smaller.

\begin{figure}[t!]
\plotfiddle{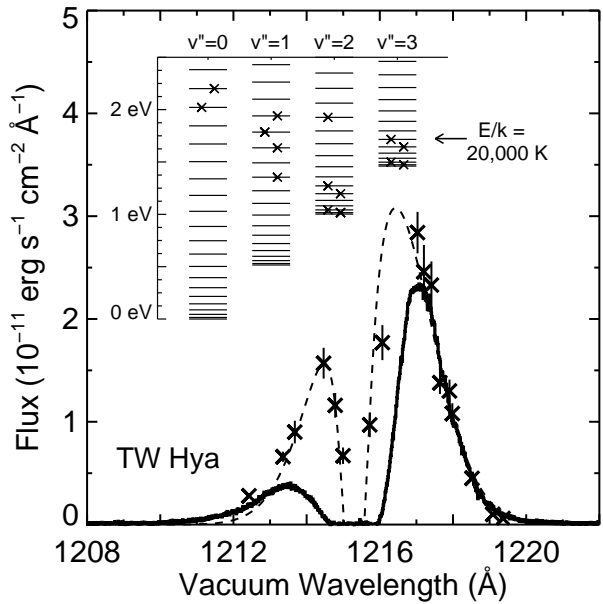}{0.05in}{0.}{230}{230}{2}{0}
\vspace{-0.1in}
\caption{\small\baselineskip 10pt
\Lya\ emission from TW Hya, an accreting T Tauri star,
and a reconstruction of the \Lya\ profile seen by the circumstellar \Htwo.
Each observed \Htwo\ progression (with a common excited state) yields a
single point in the reconstructed \Lya\ profile. The wavelength of
each point in the reconstructed \Lya\ profile corresponds to the
wavelength of the upward transition that pumps the progression.
The required excitation energies for the \Htwo\ \textit{before} the
pumping is indicated in the inset energy level diagram. There are
no low excitation states of \Htwo\ with strong transitions that overlap
\Lya. Thus, the \Htwo\ must be very warm to be preconditioned for
pumping and subsequent fluorescence.} 
\end{figure}

% Only a tiny fraction of possible FUV H2 lines are observed

Considering only radiative transitions with spontaneous rates above
10$^7$ s$^{-1}$, \Htwo\ has about 9000 possible Lyman-band (B-X)
transitions from 850-1650 \AA\ and about 5000 possible Werner-band
(C-X) transitions from 850-1300 \AA\ ({\em Abgrall et al.}, 1993a,b). However,
only about 200 FUV transitions have actually been detected in spectra
of accreting TTS.
%
% FUV H2 lines are fluoresced
%
Detected \Htwo\ emission lines in the FUV all originate from about two
dozen radiatively pumped states, each more than 11 eV above ground.
These pumped states of \Htwo\ are the only ones connected to the ground
electronic configuration by strong radiative transitions that overlap
the broad \Lya\ emission that is characteristic of accreting TTS (see
Fig.\ 4). Evidently, absorption of broad \Lya\ emission pumps the \Htwo\
fluorescence.
%
% The H2 must be initially excited before fluorescence can occur.
%
The two dozen strong \Htwo\ transitions that happen to overlap the broad
\Lya\ emission are all pumped out of high $v$ and/or high $J$ states at
least 1 eV above ground (see inset in Fig.~4). This means some
mechanism must excite \Htwo\ in the ground electronic configuration,
before \Lya\ pumping can be effective. If the excitation mechanism is
thermal, then the gas must be roughly 10$^3$ K to obtain a significant
\Htwo\ population in excited states. 

% H2 emission is ubiquitous in accreting TTS

\Htwo\ emission is a ubiquitous feature of accreting TTS. Fluoresced \Htwo\
is detected in the spectra of 22 out of 24 accreting TTS observed in the
FUV by HST/STIS ({\em Herczeg et al.}, 2002; {\em Walter et al.}, 2003; 
{\em Calvet et al.}, 2004; {\em Bergin et al.}, 2004; 
{\em Gizis et al.}, 2005; {\em Herczeg et al.}, 2005; 
unpublished archival data). Similarly, \Htwo\ is detected in
all 8 accreting TTS observed by HST/GHRS ({\em Ardila et al.}, 2002) and all
4 published FUSE spectra ({\em Wilkinson et al.}, 2002; 
{\em Herczeg et al.}, 2002; 2004; 2005). 
Fluoresced \Htwo\ was even detected in 13 out of 39
accreting TTS observed by IUE, despite poor sensitivity 
({\em Brown et al.}, 1981; {\em Valenti et al.,} 2000).
%
% No evidence of H2 around nonaccreting TTS, and yet it may exist
%
Fluoresced \Htwo\ has not been detected in FUV spectra of 
non-accreting TTS, despite observations of 14 stars with STIS 
({\em Calvet et al.}, 2004; unpublished archival data), 
1 star with GHRS ({\em Ardila et al.}, 2002), and 19 stars with IUE 
({\em Valenti et al.}, 2000). However,
the existing observations are not sensitive enough to prove that the
circumstellar \Htwo\ column decreases contemporaneously with the
dust continuum of the inner disk.
%
% Fluorescence becomes less efficient once accretion stops
%
When accretion onto the stellar surface stops, fluorescent pumping
becomes less efficient because the strength and breadth of \Lya\
decreases significantly and the \Htwo\ excitation needed to prime the
pumping mechanism may become less efficient. COS, if installed on
HST, will have the sensitivity to set interesting limits on \Htwo\
around non-accreting TTS in the TW Hya association.

% H2 fluxes constrain Lya profile seen by the H2

The intrinsic \Lya\ profile of a TTS is not observable at Earth, except
possibly in the far wings, due to absorption by neutral hydrogen along
the line of sight. However, observations of \Htwo\ line fluxes constrain
the \Lya\ profile seen by the fluoresced \Htwo. The rate at which a
particular \Htwo\ upward transition absorbs \Lya\ photons is equal to the
total rate of observed downward transitions out of the pumped state,
corrected for missing lines, dissociation losses, and propagation
losses. If the total number of excited \Htwo\ molecules before pumping is
known (e.g., by assuming a temperature), then the inferred
pumping rate yields a \Lya\ flux point at the wavelength of each
pumping transition (Fig.\ 4).

% Smoothness of reconstructed \Lya\ supports thermal excitation

{\em Herczeg et al.}\ (2004) applied this type of analysis to TW Hya,
treating the circumstellar \Htwo\ as an isothermal, self-absorbing slab.
Fig.~4 shows reconstructed \Lya\ flux points for the upward pumping
transitions, assuming the fluoresced \Htwo\ is at 2500 K. The smoothness
of the reconstructed \Lya\ flux points implies that the \Htwo\ level
populations are consistent with thermal excitation. Assuming an \Htwo\
temperature warmer or cooler by a few hundred degrees leads to
unrealistic discontinuities in the reconstructed \Lya\ flux points.
The reconstructed \Lya\ profile has a narrow absorption component
that is blueshifted by $-90~ \kms$, presumably due to an intervening
flow.

% H2 is unresolved in TW Hya, but extended in T Tau

The spatial morphology of fluoresced \Htwo\ around TTS is diverse.
{\em Herczeg et al.} (2002) used STIS to observe TW Hya with 50 mas
angular resolution, corresponding to a spatial resolution of 2.8 AU at
a distance of 56 pc, finding no evidence that the fluoresced \Htwo\ is
extended. At the other extreme, {\em Walter et al.} (2003) detected
fluoresced \Htwo\ up to 9 arcsec from T Tau N, but only in progressions
pumped by \Htwo\ transitions near the core of \Lya.
%
% Velocity shifts of H2 suggest both disk and outflow components
%
Fluoresced \Htwo\ lines have a main velocity component at or near the
stellar radial velocity and perhaps a weaker component that is
blueshifted by tens of $\kms$ ({\em Herczeg et al.}, 2006). These two
components are attributed to the disk and the outflow, respectively.
TW Hya has \Htwo\ lines with no net velocity shift, consistent with
formation in the face-on disk ({\em Herczeg et al.}, 2002). On the other
hand, RU Lup has \Htwo\ lines that are blueshifted by $12~ \kms$, suggesting
formation in an outflow. In both of these stars, absorption in
the blue wing of the \ion{C}{2} 1335 \AA\ wind feature strongly
attenuates \Htwo\ lines that happen to overlap in wavelength, so in
either case \Htwo\ forms inside the warm atomic wind 
({\em Herczeg et al.}, 2002; 2005).

% Velocity widths of H2 lines have weak dependence on inclination

The velocity widths of fluoresced \Htwo\ lines (after removing
instrumental broadening) range from $18~ \kms$ to $28~ \kms$ for the 7
accreting TTS observed at high spectral resolution with STIS 
({\em Herczeg et al.}, 2006). 
Line width does not correlate well with inclination.
For example, TW Hya (nearly face-on disk) and DF Tau (nearly edge-on
disk) both have line widths of $18~ \kms$. Thermal broadening is
negligible, even at 2000 K. Keplerian motion, enforced corotation, and
outflow may all contribute to \Htwo\ line width in different systems.
More data are needed to understand how velocity widths (and shifts)
depend on disk inclination, accretion rate, and other factors.

\bigskip
\noindent
\textbf{2.5 Infrared Transitions of Molecular Hydrogen} 
\bigskip
%\subsection{Infrared Transitions of Molecular Hydrogen} 

Transitions of molecular hydrogen have also been studied at longer
wavelengths, in the near- and mid-infrared.  
The $v$=1--0 S(1) transition of H$_2$ (at $2\mu$m) has been detected in
emission in a small sample of classical T Tauri stars (CTTS) and one weak
T Tauri star (WTTS; {\em Bary et al.}, 2003 and references therein). 
The narrow emission lines ($\lesssim$$10\kmps$), if arising in a disk, 
indicate an origin at
large radii, probably beyond 10\,AU.  The high temperatures
required to excite these transitions thermally (1000s\,K), in contrast
to the low temperatures expected for the outer disk, suggest that
the emission is non-thermally excited, possibly by X-rays ({\em Bary et
al.}, 2003).  
The measurement of other rovibrational transitions of H$_2$ is needed
to confirm this.

The gas mass detectable by this technique depends
on the depth to which the exciting radiation can penetrate the disk.
Thus, the emission strength may be limited
either by the strength of the radiation field, if the gas column
density is high, or by the mass of gas present, if the gas column
density is low.
While it is therefore difficult to measure total gas masses with 
this approach, clearly non-thermal processes can light 
up cold gas, making it easier to detect.  

Emission from a WTTS is surprising
since WTTS are thought to be largely devoid of circumstellar 
dust and gas, given 
the lack of infrared excesses and the low accretion rates for these
systems.  The Bary et al.\ results call this
assumption into question and suggest that longer lived gaseous
reservoirs may be present in systems with low accretion rates.  We
return to this issue in Section 4.1.

At longer wavelengths, the pure rotational transitions of H$_2$ 
are of considerable interest because molecular hydrogen
carries most of the mass of the disk, and these mid-infrared
transitions are capable of probing the $\sim$100 K temperatures
that are expected for the giant planet region of the disk.
These transitions present both advantages
and challenges as probes of gaseous disks.
On the one hand, their small A-values make them sensitive, in
principle, to very large gas masses (i.e., the transitions do not
become optically thick until large gas column densities 
$N_H$=$10^{23}-10^{24}\persqcm$ 
are reached).  On the other hand, the small A-values also 
imply 
small critical densities, which allows the
possibility of contaminating emission from gas at lower densities
not associated with the disk,
including shocks in outflows and UV excitation of ambient gas.

In considering the detectability of H$_2$ emission from gaseous 
disks mixed with dust, one issue is that the dust 
continuum can become optically thick over column densities 
$N_H \ll 10^{23}-10^{24}\persqcm$.  
Therefore, in a disk that is optically thick in the continuum 
(i.e., in CTTS), H$_2$ emission may probe smaller 
column densities. 
In this case, the line-to-continuum contrast may be low unless there  
is a strong temperature inversion in the disk atmosphere, and 
high signal-to-noise observations may be required to detect the 
emission. 
In comparison, in disk systems that are optically thin 
in the continuum (e.g., WTTS), H$_2$ could 
be a powerful probe as long as there are sufficient heating 
mechanisms (e.g., beyond gas-grain coupling) to heat the H$_2$.

A thought-provoking result from ISO was the report of approximately
Jupiter-masses of warm gas residing in $\sim$20\,Myr old debris
disk systems ({\em Thi et al.}, 2001) based on the detection of the 
28~$\mu$m and 17~$\mu$m lines of H$_2$.  This result was surprising 
because of the advanced age of the sources in which the emission 
was detected; gaseous reservoirs are expected to dissipate on 
much shorter timescales (Section 4.1).  
This intriguing result is, thus far, unconfirmed by either 
ground-based studies ({\em Richter et al.}, 2002; 
{\em Sheret et al.}, 2003; 
{\em Sako et al.}, 2005) or studies with Spitzer 
(e.g., {\em Chen et al.}, 2004).

Nevertheless, ground-based studies have detected pure rotational 
H$_2$ emission from some sources.  Detections to date include AB Aur 
({\em Richter et al.}, 2002).  
The narrow width of the emission in AB Aur ($\sim$$10~\kms$ FWHM), 
if arising in a disk, locates the emission 
beyond the giant planet region. 
Thus, an important future direction for these studies is to search 
for H$_2$ emission in a larger number of sources and 
at higher velocities, in the giant planet region of the disk.  
High resolution mid-IR spectrographs on $>$3-m telescopes 
will provide the greater sensitivity needed for such studies.

\bigskip
\noindent
\textbf{2.6 Potential Disk Diagnostics} 
\bigskip
%\subsection{Potential Disk Diagnostics} 

In a recent development, {\em Acke et al.}\ (2005) have reported
high resolution spectroscopy of the [OI]\,6300\,\AA\ line in 
Herbig AeBe stars.  The majority of the sources show a
narrow ($<$$50~\kms$ FWHM), fairly symmetric emission component
centered at the radial velocity of the star.  In some cases,
double-peaked lines are detected.  These features are interpreted
as arising in a surface layer of the disk that is irradiated by the star. 
UV photons incident on the disk surface are thought to photodissociate 
OH and H$_2$O, producing a non-thermal population of excited 
neutral oxygen that decays radiatively, producing the observed emission lines. 
Fractional OH abundances of $\sim$$10^{-7}-10^{-6}$ are needed 
to account for the observed line luminosities.

Another recent development is the report of strong absorption in the 
rovibrational bands of C$_2$H$_2$, HCN, and CO$_2$ in the 13--15~$\mu$m 
spectrum of a low-mass class I source in Ophiuchus, IRS\,46  
({\em Lahuis et al.}, 2006).  The high excitation temperature of 
the absorbing gas (400-900\,K) suggests an origin close
to the star, an interpretation that is consistent with millimeter 
observations of HCN which indicate a source size $\ll$100\,AU.
Surprisingly, high dispersion observations of rovibrational CO
(4.7~$\mu$m) and HCN (3.0~$\mu$m) show that the molecular absorption 
is {\it blueshifted} relative to the molecular cloud.  If IRS\,46
is similarly blueshifted relative to the cloud, the absorption may
arise in the atmosphere of a nearly edge-on disk.  A disk origin 
for the absorption is consistent with the observed relative abundances 
of C$_2$H$_2$, HCN, and CO$_2$ ($10^{-6}$--$10^{-5}$), which are
close to those predicted by {\em Markwick et al.}\ (2002) for
the inner region of gaseous disks ($\lesssim$2\,AU; see Section 3).
Alternatively, if IRS\,46 has approximately the same velocity as
the cloud, then the absorbing gas is blueshifted with respect to the star 
and the absorption may arise in an outflowing wind.  
Winds launched from the disk, at AU distances, may have 
molecular abundances similar to those observed if the chemical 
properties of the wind change slowly as the wind is launched.  
Detailed calculations of the chemistry of disk winds are needed to 
explore this possibility.  The molecular abundances in the 
inner disk midplane (Section 3.3) provide the initial conditions for such 
studies.

\section{\textbf{THERMAL-CHEMICAL MODELING}}

\bigskip
\noindent
\textbf{3.1 General Discussion}
\bigskip

The results discussed in the previous section illustrate the growing
potential for observations to probe gaseous inner disks.  While,
as already indicated, some conclusions can be drawn directly from
the data coupled with simple spectral synthesis modeling, harnessing
the full diagnostic potential of the observations will likely rely
on detailed models of the thermal-chemical structure 
(and dynamics) of disks.  Fortunately, the development of 
such 
models has been an active area of recent research.
Although much of the effort has been devoted to understanding the
outer regions of disks ($\sim$100\,AU; e.g., {\em Langer et al.}, 2000;
chapters by {\em Bergin et al.}\ and {\em Dullemond et al.}), recent 
work has begun to focus on the region within 10\,AU.

Because disks are intrinsically complex structures, the models
include a wide array of processes.  These encompass heating sources
such as stellar irradiation (including far UV and X-rays) and viscous
accretion; chemical processes such as photochemistry and grain
surface reactions; and mass transport via magnetocentrifugal winds,
surface evaporative flows, turbulent mixing, and accretion onto the
star.  The basic goal of the models is to calculate the density,
temperature, and chemical abundance structures that result from
these processes.  Ideally, the calculation would be fully
self-consistent, although approximations are made to simplify the
problem.

A common simplification is to adopt a specified density distribution 
and then solve the rate equations that define the chemical model.  
This is justifiable where the thermal and chemical
timescales are short compared to the dynamical timescale.
A popular choice is the $\alpha$-disk model ({\em Shakura and Sunyaev}, 1973; 
{\em Lynden-Bell and Pringle}, 1974) in which a 
phenomenological parameter $\alpha$ characterizes the efficiency of 
angular momentum transport; its vertically averaged 
value is estimated to be $\sim$$10^{-2}$ for 
T Tauri disks on the basis of
measured accretion rates ({\em Hartmann et al.}, 1998). 
Both vertically isothermal $\alpha$-disk models  
and the Hayashi minimum mass solar nebula 
(e.g., {\em Aikawa et al.}, 1999) 
were adopted in early studies. 

An improved method removes the assumption of vertical 
isothermality and calculates the vertical thermal structure 
of the disk 
including viscous accretion heating at the midplane 
(specified by $\alpha$) 
and stellar radiative heating
under the assumption that the gas and dust temperatures are the
same ({\em Calvet et al.}, 1991; {\em D'Alessio et al.}, 1999).
Several chemical models have been built using the D'Alessio density
distribution (e.g., {\em Aikawa and Herbst}, 1999; GNI04; 
{\em Jonkheid et al.}, 2004).

Starting about
2001, theoretical models showed that the gas temperature can become
much larger than the dust temperature in the 
atmospheres of outer 
({\em Kamp and van Zadelhoff}, 2001) 
and inner ({\em Glassgold and Najita}, 2001) disks.
This suggested the need to treat the gas and dust as two independent
but coupled thermodynamic systems. 
As an example of this approach, {\em Gorti and Hollenbach} (2004) 
have iteratively solved a system of chemical rate equations 
along with the equations of hydrostatic equilibrium
and thermal balance for both the gas and the dust. 

The chemical models developed so far 
are characterized by diversity as well as uncertainty.  
There is diversity in 
the adopted density distribution and external radiation field 
(UV, X-rays, and cosmic rays; the relative importance of these 
depends on the evolutionary stage) and in 
the thermal and chemical processes considered.
The relevant heating processes are less well understood than 
line cooling. 
One issue is how UV, X-ray, and cosmic rays heat the gas.
Another is the role of mechanical heating associated
with various flows in the disk, especially accretion 
(GNI04). 
The chemical processes are also less certain. 
Our understanding of astrochemistry is based mainly
on the interstellar medium, where densities and temperatures
are low compared to those of inner disks,
except perhaps in shocks and photon-dominated regions. 
New reaction pathways or 
processes may be important at the higher densities 
($> 10^{7}\percc$)
and higher radiation fields of inner disks.  
A basic challenge is to understand the thermal-chemical 
role of dust grains and PAHs.  Indeed, perhaps the most significant 
difference between models is the treatment of grain chemistry. 
The more sophisticated models include adsorption of gas onto 
grains in cold regions and desorption in warm  regions. 
Yet another level of complexity is introduced 
by transport processes which can affect the
chemistry through vertical or radial mixing. 

An important practical issue in thermal-chemical modeling is that 
self-consistent calculations become increasingly
difficult as the density, temperature, and the number of species
increase. Almost all models employ truncated chemistries with 
with somewhere from 25 to 215 species, compared with 396 in the UMIST data
base ({\em Le Teuff et al.}, 2000). The truncation
process is arbitrary, determined largely by the goals of the
calculations. {\em Wiebe et al.}, (2003) have an
objective method for selecting the most important reactions from
large data bases.  
Additional insights 
into disk chemistry are offered in the chapter by {\em Bergin et al.}

\bigskip
\noindent
\textbf{3.2 The Disk Atmosphere}
\bigskip

As noted above, {\em Kamp and van Zadelhoff} (2001) concluded 
in their model of debris disks that 
the gas and dust temperature can differ, 
as did {\em Glassgold and Najita} (2001) for T Tauri disks. The
former authors developed a comprehensive thermal-chemical model where
the heating is primarily from the dissipation of the drift velocity
of the dust through the gas. For T Tauri disks, stellar X-rays,
known to be a universal property of low-mass YSOs,
heat the gas to temperatures 
thousands of degrees  
hotter than the dust temperature.

\begin{figure}[t!]
\vspace{-0.2in}
\plotfiddle{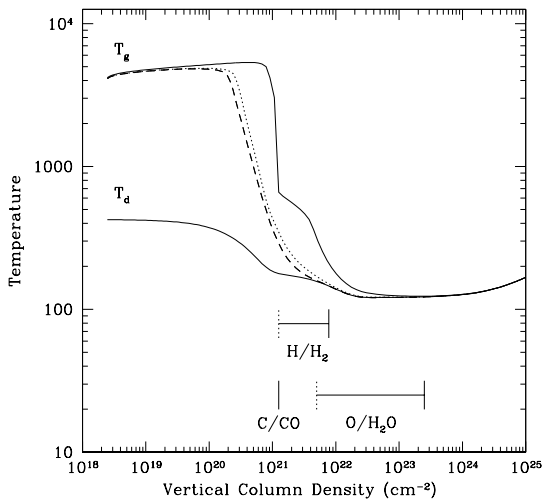}{0.05in}{0.}{220}{220}{10}{0}
\vspace{-0.3in}
\caption{\small\baselineskip 10pt
Temperature profiles from GNI04 for a protoplanetary disk atmosphere. 
The lower solid line shows the dust 
temperature of {\em D'Alessio et al.} (1999) 
at a radius of 1\,AU and a mass accretion rate of 
$10^{-8}M_\odot$ yr$^{-1}$.  
The upper curves show the corresponding gas temperature as a function of the 
phenomenological mechanical heating parameter defined by
Equation 1, $\alpha_h$ = 1 (solid line), 0.1 (dotted line), and
0.01 (dashed line).  The $\alpha_h = 0.01$ curve closely follows
the limiting case of pure X-ray heating.  The lower vertical lines
indicate the major chemical transitions, specifically CO forming
at $\sim 10^{21} {\rm cm}^{-2}$, H$_2$ forming
at $\sim 6\times 10^{21} {\rm cm}^{-2}$, and
water forming at higher column densities.} 
\end{figure}

Fig.~5 shows the vertical temperature profile obtained by {\em
Glassgold et al.}\ (2004) with a thermal-chemical model
based on the dust model of {\em D'Alessio et al.}\ (1999) for a generic
T Tauri disk. Near the midplane, the densities are high enough
to strongly couple the dust and gas.  
At higher altitudes,
the disk becomes optically thin to the stellar optical
and infrared radiation, and the temperature of the (small) grains
rises, as does the still closely-coupled gas temperature. However,
at still higher altitudes, the gas responds strongly to the 
less attenuated X-ray flux, and its temperature
rises much above the dust temperature. 
The presence of a hot X-ray heated layer above a cold midplane
layer was obtained independently by {\em Alexander et al.} (2004).

GNI04 also considered 
the possibility that the surface layers of protoplanetary disks 
are heated by the dissipation of mechanical energy. 
This might arise through 
the interaction of a wind with the upper layers of the disk 
or through disk angular momentum transport. 
Since the theoretical understanding of such processes is incomplete, 
a phenomenological
treatment is required. 
In the case of angular momentum transport, 
the most widely accepted mechanism
is the MRI ({\em Balbus and Hawley}, 1991; 
{\em Stone et al.}, 2000), which leads to the local
heating formula,
\be
\label{accheat}
\Gamma_{\rm acc} = \frac{9}{4} \alpha_h \rho c^2 \Omega, 
\ee 
where $\rho$ is the mass density, $c$ is the isothermal sound speed,
$\Omega$ is the angular rotation speed, and $\alpha_h$ is a
phenomenological parameter 
that depends on how the turbulence dissipates.
One can argue, on the basis of simulations by {\em
Miller and Stone} (2000), that midplane turbulence generates Alfv\'en
waves which, on reaching the diffuse surface regions, produce shocks
and heating. Wind-disk heating can be represented by a similar
expression on the basis of dimensional arguments.
Equation 1 is essentially an adaptation of the expression for 
volumetric heating in an $\alpha$-disk model, where $\alpha$ can in 
general depend on position. GNI04 used the notation $\alpha_h$ 
to distinguish its value in the disk atmosphere from the usual 
midplane value.

In the top layers fully exposed to X-rays, the gas temperature
at 1\,AU is $\sim$5000\,K. Further down, there is a 
warm transition region (500--2000\,K) composed mainly of atomic 
hydrogen but with carbon fully associated into CO. 
The conversion from atomic H to H$_2$ is reached at a column 
density of $\sim$$6\times 10^{21}\persqcm$, with more complex 
molecules such as water forming deeper in the disk. 
The location and thickness of the warm molecular region 
depends on the strength of the surface heating. The curves 
in Fig.~5 illustrate this dependence for a T Tauri disk at
$r=1$\,AU.  With $\alpha_h = 0.01$, X-ray heating dominates this
region, whereas with $\alpha_h > 0.1$, mechanical heating dominates.

Gas temperature inversions can also be produced by UV radiation
operating on small dust grains and PAHs, as demonstrated by
the thermal-chemical models of 
{\em Jonkheid et al.} (2004) 
and {\em Kamp and Dullemond} (2004). {\em
Jonkheid et al.}\ use the {\em D'Alessio et al.}\ (1999) model and
focus on the disk beyond 50\,AU. At this radius, the gas temperature
can rise to 800\,K or 200\,K, depending on whether small grains
are well mixed or settled. For a thin disk and a high stellar UV
flux, {\em Kamp and Dullemond} obtain temperatures that asymptote
to several 1000\,K inside 50\,AU. 
Of course these results are
subject to various assumptions that have been made about the stellar
UV, the abundance of PAHs, and the growth and settling of dust
grains.

Many of the earlier chemical models, oriented towards outer
disks (e.g., {\em Willacy and Langer}, 2000; {\em Aikawa
and Herbst}, 1999; 2001; {\em Markwick et al.}, 2002), adopt a value
for the {\it stellar} UV radiation field that is $10^4$ times larger
than Galactic 
at a distance of 100\,AU. This choice can be traced
back to early IUE measurements of the stellar UV beyond 1400\,\AA\, 
for several TTS ({\em Herbig and Goodrich}, 1986).  Although
the UV flux from TTS covers a range of values and is
undoubtedly time-variable, detailed studies with IUE
(e.g., {\em Valenti et al.}, 2000; {\em Johns-Krull et al.}, 2000)
and FUSE (e.g., {\em Wilkinson et al.}, 2002; {\em Bergin et al.}, 
2003) indicate that it decreases into the FUV domain with
a typical value $\sim$$10^{-15} \erg \, \psqcm \ps$\,\AA $^{-1}$,
much smaller than earlier estimates.
A flux of $\sim$$10^{-15}\erg \, \psqcm \ps$\,\AA $^{-1}$ at Earth
translates into a value at 100\,AU of $\sim$100 times the traditional 
Habing value for the interstellar medium.
The data in the FUV range are sparse, unfortunately, as a function 
of age or the evolutionary state of the system.  More measurements of 
this kind are needed since 
it is obviously important to use realistic fluxes in the crucial FUV band
between 912 and 1100\,\AA\ where atomic C can be photoionized and
H$_2$ and CO photodissociated 
({\em Bergin et al.}, 2003 and the chapter by {\em Bergin et al.}). 

Whether stellar FUV or X-ray radiation dominates the ionization, 
chemistry, and heating of protoplanetary disks is important because of 
the vast difference in photon energy. 
The most direct physical consequence
is that FUV photons cannot ionize H, and thus the abundance of
carbon provides an upper limit to the ionization level produced by
the photoionization of heavy atoms, 
$x_e\sim$$10^{-4}$--$10^{-3}$. Next,
FUV photons are absorbed much more readily than X-rays, although
this depends on the size and spatial distribution of the dust grains,
i.e, on grain growth and sedimentation. Using realistic numbers for
the FUV and X-ray luminosities of TTS, we estimate that
$L_{\rm FUV} \sim \LX$. The rates used in many  early chemical
models correspond to $\LX \ll L_{\rm FUV} $.
This suggests that future chemical modeling of protoplanetary disks 
should consider both X-rays and FUV in their treatment of ionization,
heating, and chemistry.

\bigskip
\noindent
\textbf{3.3 The Midplane Region}
\bigskip

Unlike the warm upper atmosphere of the disk, which is accessible
to observation,  
the optically thick midplane is much more difficult to
study. Nonetheless, it is extremely important for understanding the dynamics
of the basic flows in star formation such as accretion and outflow. 
The important role of the
ionization level for disk accretion via the MRI was pointed out by
{\em Gammie} (1996). The physical reason is that collisional coupling
between electrons and neutrals is required to transfer the turbulence
in the magnetic field to the neutral material of the disk. Gammie
found that Galactic cosmic rays cannot penetrate beyond a 
surface layer of the disk. He suggested that accretion only occurs
in the surface of the inner disk (the ``active region'') and not in
the much thicker midplane region (the ``dead zone'') where the
ionization level is too small to mediate the MRI.  

{\em Glassgold et al.}\ (1997) argued that the Galactic cosmic rays never
reach the inner disk because they are blown away by the stellar
wind, much as the solar wind excludes Galactic cosmic rays.  They
showed that YSO X-rays do almost as good a job as cosmic rays in
ionizing surface regions, thus preserving the layered accretion
model of the MRI for YSOs. {\em Igea and Glassgold} (1999) supported
this conclusion with a Monte Carlo calculation of X-ray transport
through disks, demonstrating that scattering plays an important
role in the MRI by extending the active surface layer to column
densities greater than $10^{25} \, \psqcm$, approaching the
Galactic cosmic ray range used by {\em Gammie} (1996).
This early work showed that the theory of disk ionization and
chemistry is crucial for understanding the role of the MRI for YSO
disk accretion and possibly for planet formation. Indeed, {\em Glassgold,
Najita, and Igea} suggested that Gammie's dead zone might provide
a good environment for the formation of planets. 

These challenges
have been taken up by several groups (e.g., {\em Sano et al.}, 2000;
{\em Fromang et al.}, 2002; {\em Semenov et al.}, 
2004; {\em Kunz and Balbus}, 2004; {\em Desch}, 2004; {\em
Matsumura and Pudritz}, 2003, 2005; and {\em Ilgner and Nelson}, 
2006a,b). {\em Fromang et al.}\ discussed many of the issues that
affect the size of the dead zone: differences in the disk model,
such as a Hayashi disk or a standard $\alpha$-disk; temporal evolution
of the disk; the role of a small abundance of heavy atoms that
recombine mainly radiatively; and the value of the magnetic Reynolds
number. {\em Sano et al.} (2000) explored the role played by small dust
grains in reducing the electron fraction when it becomes as small
as the abundance of dust grains. They showed that the dead zone 
decreases and eventually vanishes as the grain size increases or
as sedimentation towards the midplane proceeds.
More recently, {\em Inutsuka and Sano} (2005) have suggested that
a small fraction of the energy dissipated by the MRI leads to the
production of fast electrons with energies sufficient to ionize H$_2$. 
When coupled with vertical mixing of highly ionized surface regions,
Inutsuka and Sano argue that the MRI can self generate the ionization 
it needs to be operative throughout the entire disk.

Recent chemical modeling ({\em Semenov et al.}, 2004;
{\em Ilgner and Nelson}, 2006a,b) confirms that the level of ionization
in the midplane is affected by many microphysical
processes. These include the abundances of radiatively-recombining atomic
ions, molecular ions, small grains, and PAHs. The proper treatment of the
ions represents a great challenge for disk chemistry, one made
particularly difficult by the lack of observations of the dense gas at the
midplane of the inner disk. Thus the uncertainties in inner disk
chemistry preclude definitive quantitative conclusions about the
midplane ionization of  protoplanetary disks. Perhaps the biggest
wild card is the issue of grain growth, emphasized anew by {\em
Semenov et al.}, (2004). If the disk grain size
distribution were close to interstellar, then the small grains would
be effective in reducing the electron fraction and producing dead
zones. But significant grain growth is expected {\em and} observed
in the disks of YSOs, limiting the extent of dead zones (e.g.,
{\em Sano et al.}, 2002).

The broader chemical properties of the {\it inner} midplane region are
also of great interest since most of the gas in the disk is within
one or two scale heights.  
The chemical composition
of the inner midplane gas is important because it provides the
initial conditions for outflows and for the formation of planets
and other small bodies; it also determines whether the MRI operates. 
Relatively little work has been done on the midplane
chemistry of the inner disk. For example, GNI04 excluded N and S
species and restricted the carbon chemistry to species closely
related to CO. However, {\em Willacy et al.}\ (1998), {\em Markwick
et al.}\ (2002), and {\em Ilgner et al.}\ (2004) have carried out
interesting calculations that shed light on a possible rich organic
chemistry in the inner disk.

Using essentially the same chemical model, these authors follow
mass elements in time as they travel in a steady accretion
flow towards the star. At large distances, the gas is subject to
adsorption, and at small distances to thermal desorption. In between
it reacts on the surface of the dust grains; on being liberated
from the dust, it is processed by gas phase chemical reactions. The
gas and dust are assumed to have the same temperature, and all
effects of stellar radiation are ignored. The ionizing sources are
cosmic rays and $^{26}$Al. Since the collisional ionization of low 
ionization potential atoms is ignored, a very low ionization level 
results. {\em Markwick et al.}\ improve on {\em Willacy et al.}\ by
calculating the vertical variation of the temperature, and {\em
Ilgner et al.}\ consider the effects of mixing. Near 1\,AU, H$_2$O
and CO are very abundant, as predicted by simpler models, but {\em
Markwick et al.}\ find that CH$_4$ and CO have roughly equal abundances.
Nitrogen-bearing molecules, such as NH$_3$, HCN, and HNC are also predicted
to be abundant, as are a variety of hydrocarbons such as CH$_4$, C$_2$H$_2$,
C$_2$H$_3$, C$_2$H$_4$, etc. {\em Markwick et al.}\ also simulate
the presence of penetrating X-rays and find increased column densities
of CN and HCO$^+$. 
Despite many uncertainties, these predictions 
are of interest for our future understanding of the midplane region.

Infrared spectroscopic searches for hydrocarbons in disks may be 
able to test these predictions.  
For example, {\em Gibb et al.}\ (2004) searched for CH$_4$ in absorption 
toward HL Tau.  The upper limit on the abundance of 
CH$_4$ relative to CO ($<$1\%) in the absorbing gas 
may contradict the predictions of {\em Markwick et al.}\ (2002) 
if the absorption arises in the disk atmosphere. 
However, some support for the {\em Markwick et al.}\ (2002) model comes 
from a recent report by {\em Lahuis et al.}\ (2006)
of a rare detection by {\it Spitzer} of C$_2$H$_2$ and HCN in
absorption towards a YSO, with ratios close to
those predicted for the inner disk (Section 2.6). 

\bigskip
\noindent
\textbf{3.4 Modeling Implications}
\bigskip

An interesting implication of the irradiated disk atmosphere models
discussed above is that the region
of the atmosphere over which the gas and dust temperatures differ 
includes the region that is accessible
to observational study.  Indeed, the models have interesting 
implications for some of the observations presented in Section 2. 
They can account roughly 
for the unexpectedly warm gas temperatures that have been found for
the inner disk region based on the CO fundamental (Section 2.3) and UV
fluorescent H$_2$ transitions (Section 2.4).  
In essence, the warm gas temperatures arise from the direct heating 
of the gaseous component and the poor thermal coupling between the 
gas and dust components at the low densities characteristic 
of upper disk atmospheres.  
The role of X-rays in heating disk atmospheres has some support from  
the results of {\em Bergin et al.}\ (2004); they suggested that some of 
the UV H$_2$ emission from TTS arises 
from excitation by fast electrons produced by X-rays. 

In the models, CO is found to form at a column density 
$N_H$$\simeq$$10^{21}\persqcm$ 
and temperature $\sim$1000\,K in the radial range 0.5--2\,AU 
(GNI04; Fig.~5), 
conditions similar to those deduced for the emitting gas 
from the CO fundamental lines ({\em Najita et al.}, 2003). 
Moreover, CO is abundant in a region 
of the disk that is predominantly atomic hydrogen, a situation 
that is favorable for exciting the rovibrational transitions 
because of the large collisional excitation cross section 
for H + CO inelastic scattering.
Interestingly, X-ray irradiation alone is probably 
insufficient to explain the strength of the CO emission 
observed in actively-accreting TTS.  
This suggests that 
other processes may be important in heating disk atmospheres. 
GNI04 have explored the role of mechanical 
heating.  Other possible heating processes are FUV irradiation  
of grains and or PAHs. 

Molecular hydrogen column densities comparable to the UV 
fluorescent column of $\sim$$5\times 10^{18}\psqcm$ observed 
from TW Hya are reached at 1\,AU 
at a total vertical hydrogen column density
of $\sim$$5\times 10^{21}\psqcm$, where the fractional abundance 
of H$_2$ is $\sim$$10^{-3}$ (GNI04; Fig.~5). 
Since Ly$\alpha$ photons must traverse the entire 
$\sim$$5\times 10^{21}\psqcm$ in order to excite the emission, 
the line-of-sight dust opacity through this column must be relatively low.  
Observations of this kind, when combined with atmosphere models, 
may be able to constrain the gas-to-dust ratio in disk atmospheres, 
with consequent implications for grain growth and settling. 

Work in this direction has been carried out by {\em Nomura and Millar} 
(2005).  They have made a detailed thermal model of a
disk that includes the formation of H$_2$ on grains, destruction via FUV
lines, and excitation by Ly$\alpha$ photons. The gas at the surface is
heated primarily by the photoelectric effect on dust grains and PAHs,
with a dust model 
appropriate for interstellar clouds, i.e., 
one that reflects little grain growth. 
Both interstellar and stellar UV radiation
are included, the latter based on observations of TW Hya.  The gas
temperature at the surface of their flaring disk model reaches 1500\,K
at 1\,AU. They are partially successful in accounting for the
measurements of {\em Herczeg et al.}\ (2002), but their model
fluxes fall short by a factor of five or so.  A likely defect in
their model is that the calculated temperature of the disk surface is
too low, a problem that might be remedied by reducing the UV attenuation 
by dust and by including X-ray or other surface heating processes.

The relative molecular abundances that are predicted by 
these non-turbulent, layered  
model atmospheres are also of interest.  At a distance of 1\,AU, the
calculations of GNI04 indicate that the relative abundance of 
H$_2$O to CO is $\sim$$10^{-2}$ in the disk atmosphere for
column densities $<$$10^{22}\persqcm$; 
only at column densities $>$$10^{23}\persqcm$ are H$_2$O and CO 
comparably abundant.  
The abundance ratio in the atmosphere is significantly lower than 
the few relative abundances measurements to date (0.1--0.5) 
at $<$0.3\,AU 
({\em Carr et al.}, 2004; Section 2.2). 
Perhaps layered
model atmospheres, when extended to these 
small radii, will be able to account for the abundant water that is detected. 
If not, the large water abundance may be evidence of strong vertical 
(turbulent) mixing that carries abundant water from deeper in the 
disk up to the surface. 
Thus, it would be of great interest to develop the modeling for the 
sources and regions where water is observed in the context of 
both layered models and those with vertical mixing. 
Work in this direction has the potential to place unique constraints 
on the dynamical state of the disk.

\section{\textbf{CURRENT AND FUTURE DIRECTIONS}}
As described in the previous sections, significant progress has 
been made in developing both observational probes of gaseous 
inner disks as well as the theoretical models that are needed to 
interpret the observations.  In this section, we describe some  
areas of current interest as well as future directions 
for studies of gaseous inner disks.

\bigskip
\noindent
\textbf{4.1 Gas Dissipation Timescale} 
\bigskip
%\subsection{Gas Dissipation Timescale} 

The lifetime of gas in the inner disk is of interest in the context 
of both giant and terrestrial planet formation. 
Since significant gas must be present in the disk in order for a 
gas giant to form, the gas dissipation timescale in the giant 
planet region of the disk can help to identify dominant pathways 
for the formation of giant planets.  A short dissipation time scale 
favors processes such as gravitational instabilities which can 
form giant planets on short time scales ($< 1000$ yr; {\em Boss}, 1997; 
{\em Mayer et al.}, 2002). 
A longer dissipation time scale accommodates the more leisurely 
formation of planets in the core accretion scenario (few--10\,Myr; 
{\em Bodenheimer and Lin}, 2002).

Similarly, 
the outcome of terrestrial 
planet formation (the masses and eccentricities of the planets and 
their consequent habitability) may depend sensitively on   
the residual gas in the terrestrial planet region of the disk 
at ages of a few Myr. 
For example, in the picture of terrestrial planet formation 
described by {\em Kominami and Ida} (2002), 
if the gas column density in this region is $\gg$$1 \gpersqcm$  
at the epoch when protoplanets assemble to form terrestrial 
planets, 
gravitational gas drag is strong enough to circularize the orbits 
of the protoplanets, making it difficult for them to collide and 
build Earth-mass planets.  In contrast, if the gas column density is 
$\ll$$1 \gpersqcm$, Earth-mass planets can be produced, but gravitational 
gas drag is too weak to recircularize their orbits. 
As a result, only a narrow range of gas column densities around 
$\sim$$1 \gpersqcm$ is expected to lead to planets with the Earth-like 
masses and 
low eccentricities that we associate with habitability on Earth.

From an observational perspective,  
relatively little is known about the evolution of the 
gaseous component.  Disk lifetmes are typically inferred from 
infrared excesses that probe the dust component of the disk, 
although processes such as grain growth, planetesimal formation, 
and rapid grain inspiraling produced by gas drag ({\em Takeuchi and Lin}, 2005) 
can compromise dust as a tracer of the gas. 
Our understanding of disk lifetimes can be improved 
by directly probing the gas content of disks and using 
indirect probes of disk gas content such as stellar accretion rates 
(see {\em Najita,} 2006 for a review of this topic).  

Several of the diagnostics decribed in Section 2 may 
be suitable as direct probes of disk gas content. 
For example, transitions of H$_2$ and other molecules and atoms 
at mid- through far-infrared wavelengths are thought to be 
promising probes of the giant planet region of the disk 
({\em Gorti and Hollenbach},\ 2004). 
This is a important area of investigation currently for the 
Spitzer Space Telescope and, in the future, for Herschel  
and 8- to 30-m ground-based telescopes.

Studies of the lifetime of gas in the terrestrial planet region are
also in progress.  The CO transitions are well suited for this purpose
because the transitions of CO and its isotopes probe gas column
densities in the range of interest ($10^{-4}-1 \gpersqcm$).  
A current study by Najita, Carr, and Mathieu, 
which explores the residual gas content of optically thin disks 
({\em Najita}, 2004),  
illustrates some of the challenges 
in probing the residual gas content of disks.  
Firstly,
given the well-known correlation between IR excess and accretion
rate in young stars (e.g., {\em Kenyon and Hartmann}, 1995), CO emission
from sources with optically thin inner disks may be intrinsically
weak if accretion contributes significantly to heating disk
atmospheres.  
Thus, high signal-to-noise spectra may be 
needed to detect this emission. 
Secondly, since the line emission may be intrinsically weak, 
structure in the stellar photosphere may complicate the 
identification of emission features. 
Fig.~6 shows an example in which CO absorption in the stellar 
photosphere of TW Hya likely veils weak emission from the 
disk.  Correcting for the stellar photosphere would not only amplify 
the strong $v$=1--0 emission that is clearly present 
(cf. {\em Rettig et al.}, 2004), it would also uncover 
weak emission in the higher vibrational lines, confirming the 
presence of the warmer gas probed by the UV fluorescent lines 
of H$_2$ ({\em Herczeg et al.}, 2002). 

\begin{figure}[t!]
\vspace{-1.2in}
\plotfiddle{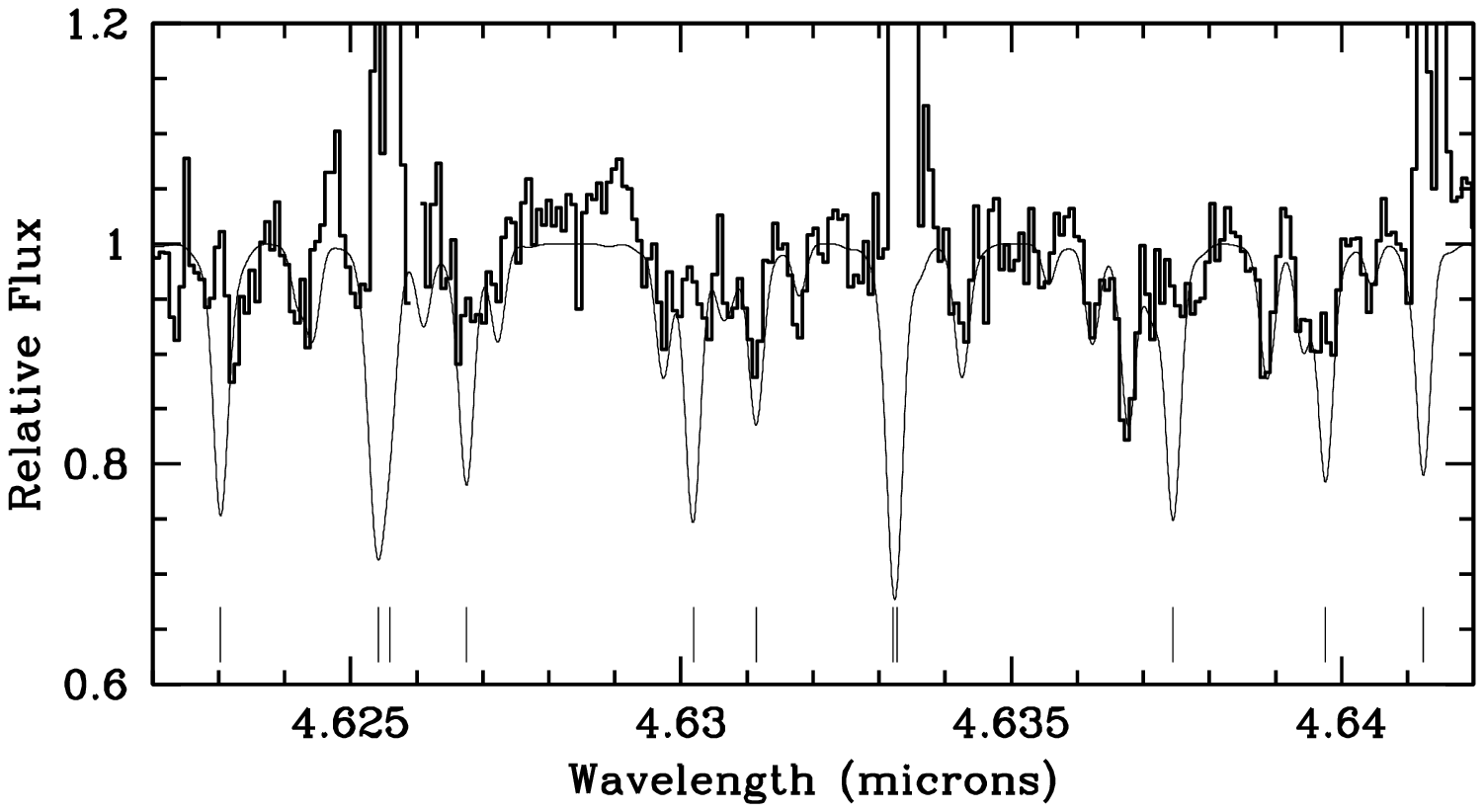}{0.05in}{0.}{300}{300}{-30}{0}
\vspace{-2.6in}
\plotfiddle{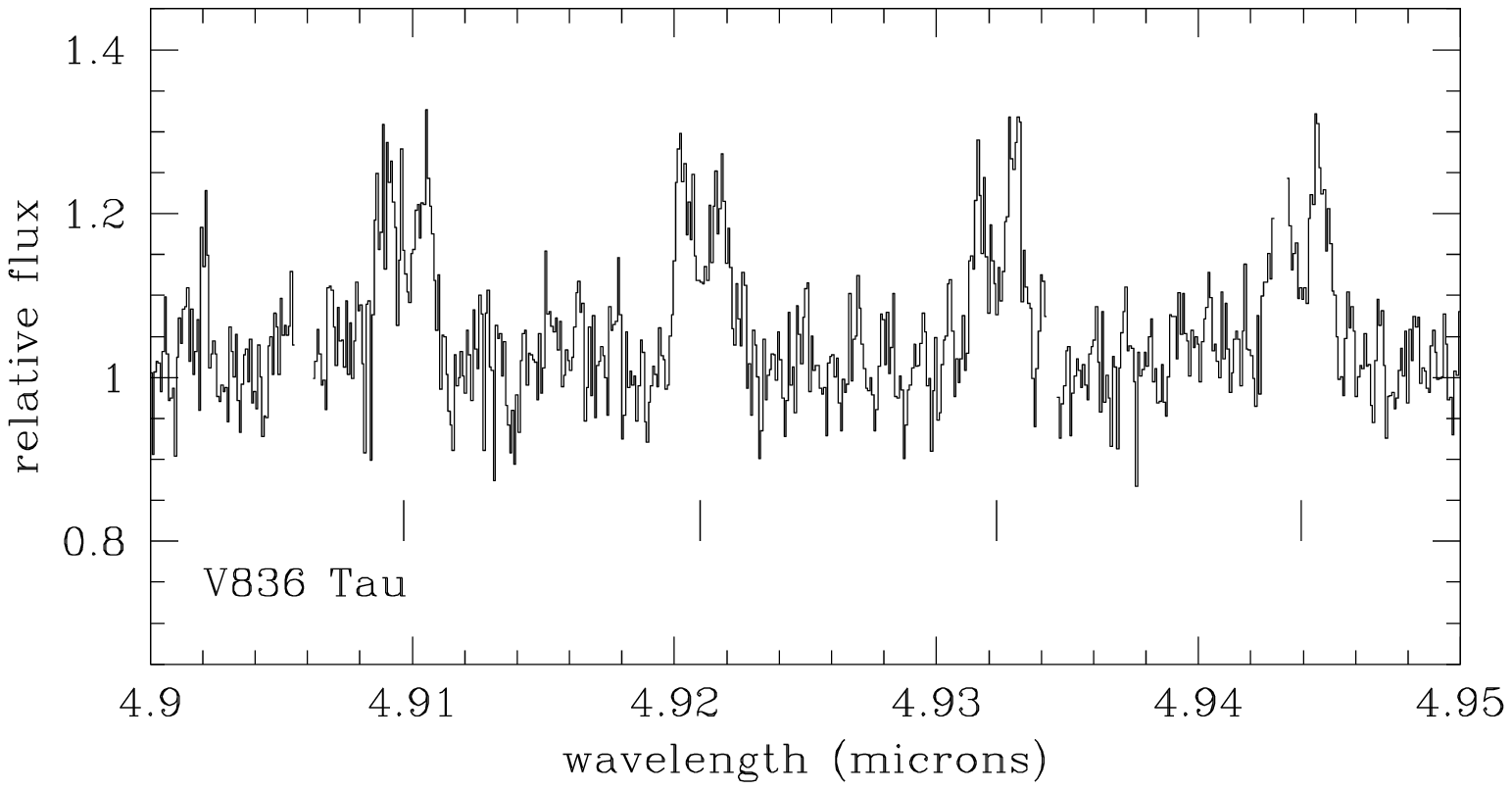}{0.05in}{0.}{300}{300}{-30}{0}
\vspace{-1.5in}
\caption{\small\baselineskip 10pt
(Top) Spectrum of the transitional disk system TW Hya at 4.6~$\micron$ 
(histogram). 
The strong emission in the $v$=1--0 CO fundamental lines extend 
above the plotted region. 
Although the model stellar photospheric spectrum (light solid line) fits 
the weaker features in the TW Hya spectrum,   
it predicts stronger absorption in the low vibrational CO 
transitions (indicated by the lower vertical lines) than is 
observed.  This suggests that the stellar 
photosphere is veiled by CO emission from warm disk gas.  
(Bottom) 
CO fundamental emission from the transitional disk system 
V836 Tau.  Vertical lines mark the approximate 
line centers at the velocity of the star.  The velocity widths 
of the lines locate the emission within a few AU of the star, 
and the relative strengths of the lines suggest optically thick 
emission.  Thus, a large reservoir of gas may be present in the 
inner disk despite the weak infrared excess from this portion 
of the disk. 
} 
\end{figure}

Stellar accretion rates 
provide a complementary probe of the gas content of inner disks. 
In a steady accretion disk, the column density $\Sigma$ is related to 
the disk accretion rate $\Mdot$ 
by a relation of the form $\Sigma\propto \Mdot/\alpha T$, where 
$T$ is the disk temperature. 
A relation of this form allows us to infer $\Sigma$ from $\Mdot$ 
given a value for the viscosity parameter $\alpha$. 
Alternatively, the relation could be calibrated empirically 
using measured disk column densities.  

Accretion rates are available for many sources in the age 
range 0.5--10\,Myr (e.g., {\em Gullbring et al.}, 1998; 
{\em Hartmann et al.}, 1998; 
{\em Muzerolle et al.}, 1998, 2000).  
A typical value of $10^{-8}\Msunperyr$ for TTS 
corresponds to a(n active) disk column density of 
$\sim$$100 \gpersqcm$ at 1\,AU for 
$\alpha$=0.01 ({\em D'Alessio et al.}, 1998). 
The accretion rates show an overall decline with time 
with a large dispersion at any given age.  
The existence of 10\,Myr old sources with accretion rates as large 
as $10^{-8}\,\Msunperyr$ ({\em Sicilia-Aguilar et al.}, 2005) suggests 
that gaseous disks may be long lived in some systems.

Even the lowest measured accretion rates may be dynamically significant.  
For a system like V836 Tau (Fig.\ 6), a $\sim$3\,Myr old 
({\em Siess et al.}, 1999) system with an optically thin inner disk, 
the stellar accretion rate of $4\times 10^{-10}\Msunperyr$ 
({\em Hartigan et al.}, 1995; {\em Gullbring et al.}, 1998) would correspond
to $\sim$$4 \gpersqcm$ at 1\,AU.  Although the accretion rate 
is irrelevant for the buildup of the stellar mass, it corresponds 
to a column density that would 
favorably impact terrestrial planet formation.  
More interesting perhaps
is St34, a TTS with a Li depletion age of 25\,Myr; 
its stellar accretion rate of $2\times 10^{-10}\Msunperyr$ 
({\em White and Hillenbrand}, 2005) 
suggests a dynamically significant reservoir of gas
in the inner disk region.  These examples suggest that dynamically
significant reservoirs of gas may persist even after inner disks 
become optically thin and over the timescales 
needed to influence the outcome of terrestrial planet formation. 

The possibility of long lived gaseous reservoirs can be confirmed 
by using the diagnostics described in Section 2 to measure total disk 
column densities.  Equally important, a measured the disk 
column density, combined with the stellar accretion rate, would allow 
us to infer a value for viscosity parameter $\alpha$ for the system.  
This would be another way of constraining the disk accretion mechanism.

\bigskip
\noindent
\textbf{4.2 Nature of Transitional Disk Systems} 
\bigskip
%\subsection{Nature of Transitional Disk Systems} 

Measurements of the gas content and distribution in inner disks 
can help us to identify systems in various states of planet 
formation.  Among the most interesting objects to study in this 
context are the transitional disk systems, which possess 
optically thin inner and optically thick outer disks.   
Examples of this class of objects include TW Hya, GM Aur, DM Tau, 
and CoKu Tau/4 ({\em Calvet et al.}, 2002; {\em Rice et al.}, 2003; 
{\em Bergin et al.}, 2004; {\em D'Alessio et al.}, 2005; 
{\em Calvet et al.}, 2005).  
It was suggested early on that optically thin inner disks might 
be produced by the dynamical sculpting of the disk by orbiting giant 
planets ({\em Skrutskie et al.}, 1990; see also {\em Marsh and Mahoney}, 1992). 

Indeed, optically thin disks may arise in multiple phases of disk 
evolution.
For example, as a first step in planet formation (via core accretion), 
grains are expected to grow into planetesimals and eventually 
rocky planetary cores, producing a region of the disk that has reduced 
continuum opacity but is gas-rich.  
These regions of the disk may therefore show strong line emission.
Determining the fraction of 
sources in this phase of evolution may help to establish the 
relative time scales for planetary core formation and the accretion of  
gaseous envelope. 

If a planetary core accretes enough gas to produce a low mass 
giant planet ($\sim$1$M_J$), it is expected to carve out a gap in its 
vicinity (e.g., {\em Takeuchi et al.}, 1996).  Gap crossing streams 
can replenish an inner disk and allow further accretion onto 
both the star and planet ({\em Lubow et al.}, 1999). 
The small solid angle subtended by the accretion streams would 
produce a deficit in the emission from both gas and dust in the 
vicinity of the planet's orbit. 
We would also expect to detect the presence of an inner disk. 
Possible examples of systems in this phase of evolution include 
GM Aur and TW Hya in which hot gas is detected close to the star as is 
accretion onto the star ({\em Bergin et al.}, 2004; {\em Herczeg et al.}, 2002; 
{\em Muzerolle et al.}, 2000).  The absence of gas in the vicinity of 
the planet's orbit would help to confirm this interpretation. 

Once the planet accretes enough mass via the accretion streams to 
reach a mass $\sim$5--10$M_J$, it is expected to cut off further accretion 
(e.g., {\em Lubow et al.}, 1999).  The inner disk will accrete onto the 
star, leaving a large inner hole and no trace of stellar accretion. 
CoKu Tau/4 is a possible example of a system in this phase of 
evolution (cf.\ {\em Quillen et al.}, 2004) since it appears to have a large 
inner hole and a low to negligible accretion rate 
($<$few\,$\times 10^{-10}\Msunperyr$). 
This interpretation predicts little gas anywhere within the orbit 
of the planet. 

At late times, 
when the disk column density around 10\,AU has decreased sufficiently 
that the outer disk is being photoevaporated away faster than 
it can resupply material to the inner disk via accretion, 
the outer disk will decouple from the inner disk, which will accrete 
onto the star, leaving an inner hole that is devoid of gas and dust 
(the ``UV Switch'' model; {\em Clarke et al.}, 2001).   
Measurements of the disk gas column density and the stellar 
accretion rate can be used to test this possibility. 
As an example, TW Hya is in the age range ($\sim$10\,Myr) where 
photoevaporation is likely to be significant. 
However, the accretion rate onto star, gas content of the inner disk 
(Sections 2 and 4), 
as well as the column density inferred for the outer disk 
($32 \gpersqcm$ at 20\,AU based on the dust SED; 
{\em Calvet et al.}, 2002) 
are all much larger than is expected in the UV switch model. 
Although this mechanism is, therefore, unlikely to explain 
the SED for TW Hya, it may explain the 
presence of inner holes in less massive disk systems of comparable age.

\bigskip
\noindent
\textbf{4.3 Turbulence in Disks} 
\bigskip
%\subsection{Turbulence in Disks} 

Future studies of gaseous inner disks may also help to 
clarify the nature of the disk accretion process.  As indicated 
in Section 2.1, evidence for suprathermal line broadening 
in disks supports the idea of a turbulent accretion process. 
A turbulent inner disk may have important 
consequences for the survival of terrestrial planets and the 
cores of giant planets.  An intriguing puzzle 
is how these objects avoid Type-I 
migration, which is expected to cause the object to lose angular 
momentum and spiral into the star on short timescales (e.g., {\em Ward}, 1997).  
A recent suggestion is that if disk accretion is turbulent, 
terrestral planets will scatter off 
turbulent fluctuations, executing a ``random walk'' which greatly 
increases the migration time as well as the chances of 
survival ({\em Nelson et al.}, 2000; see chapter by {\em Nelson et al.}). 

It would be interesting to explore this possible connection further by extending 
the approach used for the CO overtone lines to a wider 
range of diagnostics to probe the intrinsic line width as a 
function of radius and disk height.  By comparing the results 
to the detailed predictions of theoretical models, it may be 
possible to distinguish between the turbulent signature, produced 
e.g., by the MRI instability, from the turbulence that might 
be produced by, e.g., a wind blowing over the disk. 

A complementary probe of turbulence may come from exploring the
relative molecular abundances in disks.  As noted in Section 3.4,
if relative abundances cannot be explained by model
predictions for non-turbulent, layered accretion flows, a 
significant role for strong vertical mixing produced by turbulence
may be implied.  Although model-dependent, this approach toward 
diagnosing turbulent accretion appears to be less sensitive 
to confusion from wind-induced turbulence, especially if one can identify 
diagnostics that require vertical mixing from deep down in the disk. 
Another complementary approach toward probing the accretion process, 
discussed in Section 4.1, is to measure total gas column densities 
in low column density, dissipating disks in order to infer values 
for the viscosity parameter $\alpha$.

\section{\textbf{SUMMARY AND CONCLUSIONS}} 

Recent work has lent new insights on the 
structure, dynamics, and gas content of inner disks 
surrounding young stars.  
Gaseous atmospheres appear to be hotter than the dust in 
inner disks.  This is a consequence of irradiative (and 
possibly mechanical) heating of the gas as well as 
the poor thermal coupling between the gas and dust at the low 
densities of disk atmospheres.
In accreting systems, the gaseous disk appears to be turbulent 
and extends inward beyond the dust sublimation radius
to the vicinity of the corotation radius.
There is also evidence that dynamically significant reservoirs 
of gas can persist even 
after the inner disk becomes optically thin in the continuum. 
These results bear on important star and planet formation issues such as the
origin of winds, funnel flows, and the rotation rates of young
stars; the mechanism(s) responsible for disk accretion; and the role
of gas in the determining the architectures of terrestrial and
giant planets.
Although significant future work is needed to reach any conclusions 
on these issues, the future for such studies is bright.  
Increasingly detailed studies of the inner disk region should be 
possible with the advent of 
powerful spectrographs and interferometers (infrared and submillimeter) 
as well as sophisticated models that describe the coupled 
thermal, chemical, and dynamical state of the disk.

\textbf{ Acknowledgments.} We thank Stephen Strom who contributed 
significantly to the discussion on the nature of transitional disk systems. 
We also thank Fred Lahuis and Matt Richter for sharing 
manuscripts of their work in advance of publication. 
AEG acknowledges support from the NASA Origins and NSF Astronomy 
programs. 
JSC and JRN also thank the NASA Origins program for its support. 

\bigskip

\centerline\textbf{ REFERENCES }
\bigskip
\parskip=0pt
{\small
\baselineskip=11pt

% Table of the Lyman Band System of Molecular Hydrogen
\refs Abgrall H., Roueff E., Launay F., Roncin J. Y., and
Subtil, J. L. (1993) {\em Astron. Astrophys. Suppl., 101}, 273-321.

% Table of the Werner Band System of Molecular Hydrogen
\refs Abgrall H., Roueff E., Launay F., Roncin J. Y., and
Subtil J. L. (1993) {\em Astron. Astrophys. Suppl., 101}, 323-362.

\refs Acke B., van den Ancker  M. E., and Dullemond  C. P. (2005), 
\aap, 436, 209-230.

\refs Aikawa Y., Miyama S. M., Nakano T., and Umebayashi T. (1996)
{\em Astrophys. J., 467}, 684-697. 

\refs Aikawa Y., Umebayashi T., Nakano T., and Miyama S. M. (1997)
{\em Astrophys. J., 486}, L51-L54. 

\refs Aikawa Y., Umebayashi T., Nakano T., and Miyama S. M. (1999)
{\em Astrophys. J., 519}, 705-725. 

\refs Aikawa Y. and Herbst E. (1999)
{\em Astrophys. J., 526}, 314-326. 

\refs Aikawa Y. and Herbst E. (2001)
\aap, 371, 1107-1117. 

\refs Akeson R. L., Walker, C. H., Wood, K., Eisner, J. A., 
Scire, E. et al.\ (2005a) 
\apj, 622, 440-450.

\refs Akeson R. L., Boden, A. F., Monnier, J. D., Millan-Gabet, R., 
Beichman, C. et al.\ (2005b) 
\apj, 635, 1173-1181.

\refs Alexander R. D., Clarke C. J., and Pringle J. E. (2004)
\mnras, 354, 71-80.

% H2 emission in HST/GHRS spectra of BP Tau, T Tau NS, DF Tau AB,
%  DG Tau, DR Tau, RW Aur ABC, RY Tau, and RU Lup
\refs Ardila D. R., Basri G., Walter F. M., Valenti J. A., and
Johns-Krull C. M. (2002) {\em Astrophys. J., 566}, 1100-1123. 

\refs Balbus S. A. and Hawley J. F. (1991) 
{\em Astrophys. J., 376}, 214-222. 

\refs Bary J.~S., Weintraub D.~A., and Kastner J.~H. (2003) 
{\em \apj, 586}, 1138-1147.

\refs Bergin E., Calvet, N., Sitko M. L., Abgrall H., 
D'Alessio, P. et al. (2004)
{\em Astrophys. J., 614}, L133-Ll37. 

\refs Bergin E., Calvet N., D'Alessio P., and Herczeg G. J. (2003) 
{\em \apj, 591}, L159-L162. 

\refs Blake G. A. and Boogert A. C. A. (2004) 
{\em \apj, 606}, L73-L76.

\refs Blum R. D., Barbosa C. L., Damineli A., Conti P. S., 
and Ridgway S.\ (2004) {\em \apj, 617}, 1167-1176. 

\refs Bodenheimer P. and Lin D. N. C. (2002) 
{\em Ann. Rev. Earth Planet. Sci., 30} 113-148.

\refs Boss A. P. (1995)
{\em Science, 276} 1836-1839.

\refs Brittain S.~D., Rettig T.~W., Simon T., Kulesa C.,
DiSanti M.~A., and Dello Russo N. (2003)
{\em \apj, 588}, 535-544.

% First detection of H2 fluorescence in an FUV spectrum of a TTS
\refs Brown A., Jordan C., Millar T. J., Gondhalekar P., and
Wilson R. (1981) {\em Nature, 290}, 34-36.

\refs Calvet N., Patino A., Magris G., and D'Alessio P.\ (1991) 
{\em \apj, 380}, 617-630.

\refs Calvet N., D'Alessio P., Hartmann L, Wilner D., Walsh A.,
and Sitko M. (2002) 
{\em \apj, 568}, 1008-1016.

% H2 emission in HST/STIS spectra of T Tau, SU Aur, RY Tau, EZ Ori,
%  V1044 Ori, GW Ori, and CO Ori. Spectrum of P2441 is to weak.

\refs Calvet N., Muzerolle J., Brice\~no C., Hern\'andez J.,
Hartmann L., Saucedo J. L., and Gordon K. D. (2004) {\em Astron.
J., 128}, 1294-1318.

\refs Calvet N., D'Alessio P., Watson D. M., Franco-Hern\'andez R., 
Furlan, E. et al. (2005) 
{\em \apj, 630} L185-L188.

\refs Carr J. S.\ (1989) 
{\em \apj, 345}, 522-535.
 
\refs Carr J. S., Tokunaga A. T., Najita J., Shu F. H., and 
Glassgold A. E.\ (1993) 
{\em \apj, 411}, L37-L40.

\refs Carr J. S., Tokunaga A. T., and Najita J. (2004)
{\em \apj, 603}, 213-220.

\refs Chandler C. J., Carlstrom J. E., Scoville N. Z., 
Dent W. R. F., and Geballe T. R. (1993) 
{\em \apj, 412}, L71-L74.

\refs Chen C.~H., Van Cleve J.~E., Watson D.~M., Houck J.~R.,
Werner M.~W., Stapelfeldt K.~R., Fazio G.~G., and Rieke G.~H. (2004) 
{\em AAS Meeting Abstracts, 204}, 4106.

\refs Clarke C. J., Gendrin A., and Sotomayor M. (2001)
{\em MNRAS, 328}, 485-491.

\refs Colavita M., Akeson R., Wizinowich P., Shao M., 
Acton S. et al.\ (2003) 
{\em \apj, 592}, L83-L86.

\refs D'Alessio P., Canto J., Calvet N., and Lizano S. (1998)
{\em {\apj}, 500}, 411.

\refs D'Alessio P., Calvet N., and Hartmann L., Lizano S., and
Canto\'o J. (1999) 
{\em Astrophys. J., 527}, 893-909.  

\refs D'Alessio P., Calvet N., and Hartmann L. (2001) 
{\em Astrophys. J., 553}, 321-334.  

\refs D'Alessio P., Hartmann L., Calvet N., Franco-Hern\'andez R., 
Forrest W. J. et al. (2005) 
{\apj, 621}, 461-472.

\refs Desch S. (2004)
{\em Astrophys. J., 608}, 509

\refs Doppmann G. W., Greene T. P., Covey K. R., and 
Lada C. J. (2005)
{\em \apj, 130}, 1145-1170.

\refs Edwards S., Fischer W., Kwan J., Hillenbrand L., and 
Dupree A. K.\ (2003) {\em \apj, 599}, L41-L44. 

\refs Eisner J. A., Hillenbrand L. A., White R. J., Akeson R. L., 
and Sargent A. E. (2005) {\em \apj, 623}, 952-966.

\refs Feigelson E. D. and Montmerle T. (1999) 
{\em Ann. Rev. Astron. Astrophys., 37}, 363-408.  

\refs Figuer\^edo E., Blum R. D., Damineli A., and Conti P. S. (2002) 
{\em AJ, 124}, 2739-2748.
 
\refs Fromang S., Terquem C., and Balbus S. A. (2002)
{\em \mnras 339}, 19

\refs Gammie C. F. (1996) {\em Astrophys. J., 457}, 355-362. 

\refs Geballe T. R. and Persson S. E.,\ (1987) 
{\em \apj, 312}, 297-302.

\refs Gibb E. L., Rettig T., Brittain S. and Haywood R.
(2004) {\em \apj, 610}, L113-L116.

% H2 emission in HST/STIS spectra of a 10 Myr brown dwarf
\refs Gizis J. E., Shipman H. L., and Harvin J. A. (2005)
{\em Astrophys. J., 630}, L89-L91.

\refs Glassgold A. E., Najita J., and Igea J. (1997) 
{\em Astrophys. J., 480}, 344-350 (GNI97).  

\refs Glassgold A. E. and Najita J.(2001) in 
{\it Young Stars Near Earth},
ASP Conf. Ser. vol. 244 (R. Jayawardhana and T. Greene, eds.) 
pp.\ 251-255.  ASP, San Francisco.

\refs Glassgold A. E., Najita J., and Igea J. (2004) 
{\em Astrophys. J., 615}, 972-990 (GNI04). 

\refs Gorti U. and Hollenbach D. H. (2004)
{\em Astrophys. J., 613}, 424-447.

\refs Greene T. P. and Lada C. J. (1996) 
{\em AJ, 112}, 2184-2221.

\refs Gullbring E., Hartmann L., Brice\~no C., and Calvet N. (1998) 
{\em \apj, 492}, 323-341.

\refs Hanson M. M., Howarth I. D., and Conti P. S.\ (1997) 
{\em \apj, 489}, 698-718.

\refs Hartmann L. and Kenyon S.\ (1996) 
{\em Ann. Rev. Astron. Astrophys., 34}, 207-240.

\refs Hartmann L., Calvet N., Gullbring E. and D'Alessio P.
(1998) {\em \apj 495}, 385-400.

\refs Hawley J. F., Gammie C. F., and Balbus S. A. (1995) 
{\em Astrophys. J., 440}, 742-763.  

\refs Hartmann L., Hinkle K., and Calvet N.\ (2004) 
{\em \apj, 609}, 906-916.

\refs Herczeg G.~J., Linsky J.~L., Valenti J.~A., 
Johns-Krull C.~M., and Wood B.~E. (2002) 
{\em \apj, 572}, 310-325.

% {H2 emission in HST/STIS and FUSE spectra of TW Hya
\refs Herczeg G. J., Wood B. E., Linsky J. L., Valenti J. A.,
and Johns-Krull C. M. (2004) {\em Astrophys. J., 607}, 369-383. 

% H2 emission in HST/STIS spectra of RU Lup
%\refs Herczeg, G. J., Walter, F. M., Linsky, J. L., Gahm, G. F.,
%Ardila, D. R., Brown, A., Johns-Krull, C. M., Simon, M., and
%Valenti, J. A. (2005) {\em Astron. J., 129}, 2777-2791

\refs Herczeg G. J., Walter F. M., Linsky J. L., Gahm G. F., 
Ardila D. R. et al.\ (2005) {\em Astron. J., 129}, 2777-2791. 

% H2 emission in HST/STIS spectra of TW Hya, DF Tau, RU Lup, T Tau,
% DG Tau, V836 Tau, and a nondetection for V410 Tau
\refs Herczeg G. J., Linsky J. L., Walter F. M., Gahm G. F.,
and Johns-Krull C. M. (2006) {\em in preparation}

\refs Hollenbach D. J. Yorke H. W., Johnstone D. (2000)
in {\it Protostars and Planets IV}, 
(V. Mannings et al., eds.), pp. 401-428. 
Univ. of Arizona, Tucson.

\refs Ida S. and Lin D. N. C. (2004) {\em \apj, 604}, 388-413. 

\refs Igea J. and Glassgold A. E. (1999)
{\em \apj, 518}, 848-858.

\refs Ilgner M., Henning Th., Markwick A. J., and Millar T. J.  (2004)
{\em \aap, 415}, 643-659. 

\refs Ilgner M. and Nelson R. P. (2006a)
{\em \aap, 445}, 205-222.
%astro-ph/0509550.

\refs Ilgner M. and Nelson R. P. (2006b)
{\em \aap, 445}, 223-232.
%astro-ph/05099553.

\refs Inutsuka S. and Sano T. (2005) 
{\em \apj, 628}, L155-L158.

\refs Ishii M., Nagata T., Sato S., Yao Y., Jiang Z., and 
Nakaya H. (2001) 
{\em Astron. Astrophys., 121}, 3191-3206.

\refs Jonkheid B., Faas F. G. A., van Zadelhoff G.-J., and
van Dishoeck E. F. (2004) 
{\em \aap 428}, 511-521. 

\refs Johns-Krull C. M., Valenti J. A., and Linsky J. L. (2000)
{\em Astrophys. J. 539}, 815-833.

\refs Kamp I. and van Zadelhoff G.-J. (2001)
{\em \aap, 373}, 641-656. 

%\refs Kamp I., van Zadelhoff G.-J, and van Dishoeck E. (2003)
%\aap, 

\refs Kamp I. and Dullemond C. P. (2004)
{\em Astrophys. J., 615}, 991-999.

\refs Kenyon S. J. and Hartmann L. (1995) 
{\em \apjs, 101}, 117-171.

\refs Klahr H. H. and Bodenheimer P. (2003) 
{\em Astrophys. J., 582}, 869-892. 

\refs Kominami J. and Ida S. (2002) 
{\em Icarus, 157}, 43-56.

\refs Kunz M. W. and Balbus S. A. (2004)
{\em \mnras, 348}, 355-360.

\refs Lahuis F., van Dishoeck E. F., Boogert A. C. A., 
Pontoppidan K. M., Blake G. A. et al. (2006) 
{\em \apj, 636}, L145-L148. 
%astro-ph/0511786

\refs Langer W. et al.\ (2000)  
in {\it Protostars and Planets IV},
(V. Mannings et al., eds.), 
pp. 29-.  Univ. of Arizona, Tucson.

\refs Le Teuff Y., Markwick, A., and Millar, T. (2000)
{\em \aap, 146}, 157-168. 

\refs Ida S. and Lin D. N. C.\ (2004) 
{\em \apj, 604}, 388-413.

\refs Lin D. N. C., Bodenheimer P., and Richardson D. C. 
(1996) {\em Nature, 380}, 606-607. 

\refs Lubow S. H., Seibert M., and Artymowicz P. (1999)
{\em \apj, 526}, 1001-1012. 

\refs Luhman K. L., Rieke G. H., Lada C. J., and Lada E. A.  (1998) 
{\em \apj, 508}, 347-369.

\refs Lynden-Bell D. and Pringle J. E. (1974) 
{\em \mnras, 168}, 603-637. 

\refs Malbet F. and Bertout C. (1991) 
{\em \apj, 383}, 814-819. 

\refs Markwick A. J., Ilgner M., Millar T. J., and Henning Th. (2002) 
{\em \aap, 385}, 632-646.

\refs Marsh K.~A. and Mahoney M.~J. (1992) 
{\em \apj, 395}, L115-L118.

\refs Martin S. C. (1997) 
{\em \apj, 478}, L33-L36.

\refs Matsumura S. and Pudritz R. E. (2003)
{\em Astrophys. J., 598}, 645-656.

\refs Matsumura S. and Pudritz R. E. (2005)
{\em Astrophys. J., 618}, L137-L140. 

\refs Mayer L., Quinn T., Wadsley J., and Stadel J. (2002) 
{\em Science, 298}, 1756-1759.

\refs Miller K. A. and Stone J. M. (2000)
{\em \apj, 534}, 398-419.

\refs Muzerolle J., Hartmann L., and Calvet N. (1998)
{\em \aj, 116}, 2965-2974.

\refs Muzerolle J., Calvet N., Brice\~no C., Hartmann L., 
and Hillenbrand L. (2000) {\em \apj, 535}, L47-L50. 

\refs Muzerolle J., Calvet N., Hartmann L., and D'Alessio P. 
(2003) {\em \apj, 597}, L149-152. 

\refs Najita J., Carr J. S., Glassgold A. E., Shu F. H., 
and Tokunaga A. T.\ (1996) 
{\em \apj, 462}, 919-936.

\refs Najita J., Edwards S., Basri G., and Carr J. (2000)  
in {\it Protostars and Planets IV},
(V. Mannings et al., eds.), p. 457-483.
Univ. of Arizona, Tucson.

\refs Najita J., Carr J. S., and Mathieu R. D. (2003) 
{\em \apj, 589}, 931-952.

\refs Najita, J. (2004)  
in
{\it Star Formation in the Interstellar Medium} 
ASP Conf. Ser., vol.\ 323,
(D. Johnstone et al., eds.) pp. 271-277.
ASP, San Francisco.

\refs Najita, J. (2006)  
in
{\it A Decade of Extrasolar Planets Around Normal Stars} 
STScI Symposium Series, vol.\ 19,
(M. Livio, ed.) 
Cambridge U. Press, Cambridge, in press.

\refs Nelson R. P., Papaloizou J. C. B., Masset F., and 
Kley W. (2000) {\em MNRAS, 318}, 18-36. 

\refs Nomura H. and Millar T. J. (2005) 
{\em \aap, 438}, 923-938.

\refs Quillen A. C., Blackman E. G., Frank A., and 
Varni\`ere P. (2004) {\em \apj, 612}, L137-L140. 

\refs Prinn R.\ (1993) 
in {\it Protostars and Planets III}, 
(E. Levy and J. Lunine, eds.) pp. 1005-1028.
Univ. of Arizona, Tucson.

\refs Rettig T.~W., Haywood J., Simon T., Brittain S.~D., and
Gibb E. (2004)  
{\em \apj, 616}, L163-L166.

\refs Rice W. K. M., Wood K., Armitage P. J., Whitney B. A.,
and Bjorkman J. E. (2003) 
{\em MNRAS, 342}, 79-85.

\refs Richter M.~J., Jaffe D.~T., Blake G.~A., and Lacy J.~H. (2002)
{\em \apj, 572}, L161-L164.

\refs Sako S., Yamashita T., Kataza H., Miyata T., Okamoto Y. K. 
et al. (2005) 
{\em \apj, 620}, 347-354.

\refs Sano T., Miyama S., Umebayashi T., amd Nakano T. (2000)
{\em Astrophys. J., 543}, 486-501. 

\refs Scoville N., Kleinmann S. G., Hall D. N. B., and 
Ridgway S. T.\ (1983) 
{\em \apj, 275}, 201-224. 

\refs Scoville N. Z. (1985) in {\it Protostars and Planets II}, 
pp. 188-200.
Univ. of Arizona, Tucson.
 
\refs Semenov D., Widebe D., and Henning Th. (2004) 
{\em \aap, 417}, 93-106. 

\refs Shakura N. I. and Sunyaev R. A. (1973) 
{\em \aap, 24}, 337-355. 

\refs Sheret I., Ramsay Howat S. K., Dent W. R. F. (2003)
{\em MNRAS, 343}, L65-L68.

\refs Shu F. H., Johnstone D., Hollenbach D. (1993)
{\em Icarus, 106}, 92.

\refs Skrutskie M. F., Dutkevitch D., Strom S.~E.,
Edwards S., Strom K.~M., and Shure M.~A. (1990)
{\em \aj, 99}, 1187-1195.

\refs Sicilia-Aguilar A., Hartmann L.~W., Hern\'andez J.,
Brice\~no C., and Calvet N. (2005)
{\em \aj, 130}, 188-209.

\refs Siess L., Forestini M., and Bertout C. (1999)
{\em \aap, 342}, 480-491.

\refs Stone J. M., Gammie C. F., Balbus S. A., and Hawley J. F. (2000)
in {\it Protostars and Planets IV}, 
(V. Mannings et al., eds) pp. 589-611. 
Univ. Arizona, Tucson. 

\refs Strom K. M., Strom S. E., Edwards S., Cabrit S., and 
Skrutskie M. F.  (1989) 
{\em \aj, 97}, 1451-1470. 

\refs Takeuchi T., Miyama S. M., and Lin D. N. C. (1996) 
{\em \apj, 460}, 832-847.

\refs Takeuchi T., and Lin D. N. C. (2005) 
{\em \apj, 623}, 482-492.

%\refs Thi W.~F., van Dishoeck E.~F., Blake G.~A., van Zadelhoff G.~J.,
%Horn J., Becklin E.~E., Mannings V., Sargent A.~I., van den Ancker M.~E.,
%Natta A., and Kessler J. (2001)  
%{\em \apj, 561}, 1074-1094.

\refs Thi W.~F., van Dishoeck E. F., Blake G. A., van Zadelhoff G. J., 
Horn J. et al.  (2001)  
{\em \apj, 561}, 1074-1094.

\refs Thompson R.\ (1985) 
{\em \apj, 299}, L41-L44. 

\refs Trilling D. D., Lunine J. I., and Benz W. (2002) 
{\em \aap 394}, 241-251. 

\refs Valenti J. A., Johns-Krull C. M., and Linsky J. L. (2000)
{\em Astrophys. J. Suppl. 129}, 399-420.

% Spatially extended H2 emission in HST/STIS spectra of T Tau
\refs Walter F. M., Herczeg G., Brown A., Ardila D. R., Gahm
G. F., Johns-Krull C. M., Lissauer J. J., Simon M., and Valenti
J. A. (2003) {\em Astron. J., 126}, 3076-3089.

\refs Ward W. R. (1997) 
{\em Icarus, 126}, 261-281. 

\refs White R. J. and Hillenbrand L. A. (2005) 
{\em \apj, 621}, L65-L68.

\refs Wiebe D., Semenov D. and Henning Th. (2003) 
{\em \aap, 399}, 197-210. 

\refs Wilkinson E., Harper G. M., Brown A. and Herczeg G. J. (2002)
{\em \aj, 124}, 1077-1081.

\refs Willacy K. and Langer W. D. (2000)
{\em \apj, 544}, 903-920.

\refs Willacy K., Klahr H. H., Millar T. J., and Henning Th. (1998)
{\em \aap, 338}, 995-1005. 

\refs Zuckerman B., Forveille T., and Kastner J.~H. (1995) 
{\em Nature, 373}, 494-496

\end{document}